\documentclass[12pt]{article}

\usepackage{amsmath,amsfonts,enumerate,array}
\allowdisplaybreaks
\usepackage{graphicx}
\usepackage[usenames]{color}
\usepackage[english]{babel}
\usepackage{fancybox}
\usepackage{chngpage} 

\usepackage{tikz-cd}

\newcommand{\captionfonts}{\small}

\makeatletter  
\long\def\@makecaption#1#2{%
  \vskip\abovecaptionskip
  \sbox\@tempboxa{{\captionfonts #1: #2}}%
 \ifdim \wd\@tempboxa >\hsize
    {\captionfonts #1: #2\par}
  \else
    \hbox to\hsize{\hfil\box\@tempboxa\hfil}%
  \fi
  \vskip\belowcaptionskip}
\makeatother   

\usepackage{mciteplus}

\usepackage[colorlinks=true,      linkcolor=blue,      urlcolor=blue,      filecolor=blue,      citecolor=blue,
      pdfstartview=FitH,     pdfpagemode=UseNone,      bookmarksopen=true]{hyperref}  
      
\usepackage[all]{hypcap}     

\usepackage{cite}

\begin{document}

\numberwithin{equation}{section}


\mathchardef\mhyphen="2D


\newcommand{\be}{\begin{equation}} 
\newcommand{\ee}{\end{equation}} 
\newcommand{\bea}{\begin{eqnarray}\displaystyle}
\newcommand{\eea}{\end{eqnarray}}
\newcommand{\bt}{\begin{tabular}}
\newcommand{\et}{\end{tabular}}
\newcommand{\bs}{\begin{split}}
\newcommand{\es}{\end{split}}

\newcommand{\I}{\text{I}}
\newcommand{\II}{\text{II}}

\renewcommand{\a}{\alpha}	
\renewcommand{\b}{\beta}
\newcommand{\g}{\gamma}		
\newcommand{\G}{\Gamma}
\renewcommand{\d}{\delta}
\newcommand{\D}{\Delta}
\renewcommand{\c}{\chi}			
\newcommand{\C}{\Chi}
\newcommand{\p}{\psi}			
\renewcommand{\P}{\Psi}
\newcommand{\s}{\sigma}		
\renewcommand{\S}{\Sigma}
\renewcommand{\t}{\tau}		
\newcommand{\e}{\epsilon}
\newcommand{\n}{\nu}
\newcommand{\m}{\mu}
\renewcommand{\r}{\rho}
\renewcommand{\l}{\lambda}

\newcommand{\nn}{\nonumber\\} 		
\newcommand{\newotimes}{}  				
\newcommand{\diff}{\,\text{d}}		
\newcommand{\h}{{1\over2}}				
\newcommand{\Gf}[1]{\G \Big{(} #1 \Big{)}}	
\newcommand{\floor}[1]{\left\lfloor #1 \right\rfloor}
\newcommand{\ceil}[1]{\left\lceil #1 \right\rceil}

\def\cA{{\cal A}} \def\cB{{\cal B}} \def\cC{{\cal C}}
\def\cD{{\cal D}} \def\cE{{\cal E}} \def\cF{{\cal F}}
\def\cG{{\cal G}} \def\cH{{\cal H}} \def\cI{{\cal I}}
\def\cJ{{\cal J}} \def\cK{{\cal K}} \def\cL{{\cal L}}
\def\cM{{\cal M}} \def\cN{{\cal N}} \def\cO{{\cal O}}
\def\cP{{\cal P}} \def\cQ{{\cal Q}} \def\cR{{\cal R}}
\def\cS{{\cal S}} \def\cT{{\cal T}} \def\cU{{\cal U}}
\def\cV{{\cal V}} \def\cW{{\cal W}} \def\cX{{\cal X}}
\def\cY{{\cal Y}} \def\cZ{{\cal Z}}

\def\mC{\mathbb{C}} \def\mP{\mathbb{P}}  
\def\mR{\mathbb{R}} \def\mZ{\mathbb{Z}} 
\def\mT{\mathbb{T}} \def\mN{\mathbb{N}}
\def\mH{\mathbb{H}} \def\mX{\mathbb{X}}
\def\CP{\mathbb{CP}}
\def\RP{\mathbb{RP}}
\def\Z{\mathbb{Z}}
\def\N{\mathbb{N}}
\def\H{\mathbb{H}}

\newcommand{\bin}[1]{{\bf {\color{blue} BG:}} {{\color{blue}\it#1}}}
\newcommand{\shaun}[1]{{\bf {\color{red} SH:}} {{\color{red}\it#1}}}


\addtolength{\skip\footins}{0pc minus 5pt}

\def\b{\bigskip}

\begin{flushright}
\end{flushright}
\vspace{15mm}
\begin{center}
{\LARGE Four-twist effects on excitations  \\\vspace{2mm} in symmetric orbifold CFTs}
\\
\vspace{18mm}
\textbf{Bin} \textbf{Guo}{}$^{\dagger}$\footnote{guobin@csu.edu.cn}~\textbf{and} ~ \textbf{Shaun}~ \textbf{D.}~ \textbf{Hampton}{}$^{\ddagger}$\footnote{sdh2023@kias.re.kr}
\\
\vspace{8mm}
{}$^{\dagger}$~ \text{School of Physics, Central South University,}\\
	\hspace*{0.3cm} Changsha 418003, China  \\
 \vspace{8mm}
{}$^{\ddagger}$~ \text{School of Physics},\\ Korea Institute for Advanced Study,\\ Seoul 02455, Korea
\vspace{3mm}
\end{center}

\vspace{2mm}

\thispagestyle{empty}

\begin{abstract}
\vspace{2mm}

Symmetric orbifold CFTs contain twist operators that can join and split copies of the CFT. In this paper, we study the effects of four twist-2 operators on two copies of a single free boson. A recent study analyzed their effects on the vacuum, finding a nontrivial left-right mixing that arises from the fact that the covering surface is a torus, while the effects of one or two twist-2 operators do not produce such mixing. Here, we extend this analysis to excited states and find a similar left-right mixing. Furthermore, we explore the continuum, or high-energy, limit and show that the left-right mixing becomes negligible in this limit.  
 
\end{abstract}
\newpage

\setcounter{page}{1}

\numberwithin{equation}{section} 

\tableofcontents

\newpage

\section{Introduction}

Symmetric orbifold CFTs have played a significant role in understanding the AdS$_3$/CFT$_2$ correspondence \cite{Maldacena:1997re, Strominger:1996sh, Seiberg:1999xz}, providing examples of CFTs with known holographic duals. The tensionless limit of string theory \cite{Eberhardt:2018ouy,Eberhardt:2019ywk} corresponds to the free point of the symmetric orbifold CFT \cite{Seiberg:1999xz,Larsen:1999uk,Dijkgraaf:1998gf,Jevicki:1998bm}, while the supergravity limit corresponds to the strongly coupled regime. A marginal deformation moves the CFT from the free point toward the strongly coupled regime \cite{David:1999ec,Gomis:2002qi,Gava:2002xb}.
This deformation involves a twist-2 operator. Studying the effects of this twist-2 operator is essential for understanding the interpolation between the tensionless limit and the supergravity limit.

The symmetric orbifold CFT is constructed by taking multiple copies of a seed CFT and orbifolding by the permutation group. Twisted sectors occur when some copies are joined together, while in untwisted sectors, all copies remain unjoined. The twist-2 operator can join and split copies of the CFT. Its effects have been explored in various studies. For studies on a single twist-2 operator, see \cite{Avery:2010er, Avery:2010hs, Avery:2010qw, Burrington:2014yia, Carson:2014xwa, Carson:2014yxa, Guo:2022sos}, and for studies on two twist-2 operators, see \cite{Carson:2017byr, Carson:2015ohj, Carson:2016cjj}. 

One significant effect of the twist-2 operator is the creation of pairs of excitations from the vacuum. In studies with one or two twist operators, pair creation was found to occur independently in either the left-moving or right-moving sector. This means that only pairs consisting of either two left-moving modes or two right-moving modes can be created. Additionally, left-moving pair creation is a holomorphic function, while right-moving pair creation is a antiholomorphic function.
However, recent research has shown that pair creation of four twist-2 operators involves mixing between the left- and right-moving sectors \cite{Guo:2024edj}. As a result, pairs can be created with one left-moving mode and one right-moving mode. Moreover, in this case, left-moving pair creation is no longer a holomorphic function, and similarly, right-moving pair creation is no longer an antiholomorphic function. 

As shown in \cite{Guo:2024edj}, this left-right mixing comes from a torus covering map. On the torus, the left- and right-moving sectors of free fields can couple through the nontrivial periodicity. This coupling on the covering space is then transferred to the left-right mixing when four twist operators are present. In the original base space, these nontrivial periodicities appear as monodromy conditions. 

In this paper, we extend the results of \cite{Guo:2024edj}, which studied the effect of four twist-2 operators on the vacuum, to explore their effect on excited states. We follow the setup in \cite{Guo:2024edj}, where four twist-2 operators act on the untwisted sector of two copies of a free boson and ultimately return to the untwisted sector. 
Beyond pair creation, the presence of excited initial modes gives rise to two additional effects: Propagation, which describes how an initial mode transforms into a final mode, and Contraction, which accounts for the annihilation of two initial modes. We will show that these effects also exhibit left-right mixing, similar to pair creation. We also examine the continuum limit, where the energy of the excitation mode becomes very high. In this limit, the left-right mixing falls off rapidly, making it negligible at high energies. This behavior is useful for understanding the effects of multiple twist operators in the continuum limit.

The paper is organized as follows:
In Section \ref{effect of twist}, we introduce the symmetric orbifold CFT, discussing the effect of twist operators and reviewing the relevant correlator required for our computations. Sections \ref{twist effects} and \ref{twist effects contraction} provide a detailed derivation of the expressions for propagation and contraction, respectively. In Section \ref{analysis}, we analyze the properties of these effects numerically. Section \ref{limits} examines the continuum limit and large separation limits. Finally, in Section \ref{results} we collect our results, and in Section \ref{discussion} we discuss and conclude.

\section{Effect of four twist operators}\label{effect of twist}

\subsection{Symmetric product orbifold CFT}\label{orbifold review}

The symmetric product orbifold CFT is described by the target space
\be
M^N/S_N
\ee 
where $M$ denotes the target space of the seed CFT, and $S_N$ is the symmetric group acting on $N$ elements.
In this paper, we focus on the case where $N=2$ and $M = \mathbb{R}$, corresponding to two copies of a free boson. These two copies are denoted as $X^{(1)}$ and $X^{(2)}$.

The base space is a cylinder parameterized by a complex coordinate $w$
\bea\label{cylinder coord}
w=\tau + i\sigma, \quad 
-\infty < \tau < \infty, \quad 0\leq \sigma < 2\pi
\eea
We also use the complex plane coordinate $z$, related to the cylinder coordinate by $z=e^w$.

The action of $S_2$ permutes the two copies of the field, giving rise to both untwisted and twisted sectors. In the untwisted sector, the fields $X^{(1)}$ and $X^{(2)}$ are periodic
\be
X^{(i)}(\tau,\sigma+2\pi)= X^{(i)}(\tau), \qquad   i =1,2
\ee
For each copy, we define left and right moving modes at constant $\tau$ as follows 
\begin{align} \label{alpha exp}
\a^{(i)}_{n}= \int_{\sigma=0}^{2\pi}{dw\over2\pi } e^{nw}\partial_w X^{(i)}=\oint {dz\over2\pi} z^{nw}\partial_z X^{(i)}, \nn
\bar\a^{(i)}_{n}= \int_{\sigma=0}^{2\pi}{d\bar w\over2\pi } e^{n\bar w}
\partial_{\bar w} X^{(i)} = \oint {d\bar z\over2\pi} {\bar z}^{nw}\partial_{\bar z} X^{(i)}
\end{align}
where $n$ is an integer. The vacuum state $|0\rangle^{(i)}$ for each copy $i$ is defined by
\bea
\a^{(i)}_{n}|0\rangle^{(i)}=0,\quad \bar\a^{(i)}_{n}|0\rangle^{(i)}=0,\quad n\geq0
\eea
The commutation relations are given by
\bea \label{comm relation singly wound}
[\a^{(i)}_{m},\a^{(j)}_{n}]=m\d^{ij}\d_{m+n,0},\quad [\bar\a^{(i)}_{m},\bar\a^{(j)}_{n}]=m\d^{ij}\d_{m+n,0},\quad [\a^{(i)}_{m},\bar\a^{(j)}_{n}]=0
\eea
These relations can be derived from the OPEs
\begin{align} 
\partial X^{(i)}(z)\partial X^{(j)}(z')&\sim -{\d^{ij}\over (z-z')^2} + \text{regular terms}\nn
\bar \partial X^{(i)}(\bar z)\bar \partial X^{(j)}(\bar z')&\sim -{\d^{ij}\over (\bar z-\bar z')^2} + \text{regular terms}
\end{align}

There is also a twisted sector characterized by the boundary conditions
\be\label{twist bc}
X^{(1)}(\tau,\sigma+2\pi)= X^{(2)}(\tau), \qquad   X^{(2)}(\tau,\sigma+2\pi)= X^{(1)}(\tau)
\ee
To describe these two fields, it is convenient to use a single field defined on an extended base space with $0\leq \sigma<4\pi$ and periodicity $4\pi$. Since these fields and 
their mode expansions are not directly used in this paper, we refer the reader to \cite{Avery:2010er,Carson:2014ena,Guo:2023czj} for detailed discussions.

In the $z$-plane, the ground state of this twisted sector corresponds to a local operator called the twist operator, denoted as $\sigma_2$. This operator has conformal dimensions $h=\bar h =1/16$. As we focus specifically on this twist-2 operator, we omit the subscript 2 for simplicity. 

For studies on general twist operators and their correlations, we refer the reader to \cite{Lunin:2000yv,Lunin:2001pw,Pakman:2009ab,Pakman:2009mi,Pakman:2009zz,Dei:2019iym}. In the D1D5 CFT, a marginal deformation operator can be constructed using the twist-2 operator.
This marginal operator moves the D1D5 CFT from the free point towards the gravity regime and is essential for understanding anomalous dimensions\cite{Gava:2002xb,Gaberdiel:2015uca,Hampton:2018ygz,Guo:2020gxm,Benjamin:2021zkn,Lima:2021wrz,Apolo:2022fya,Guo:2022ifr,Gaberdiel:2023lco,frolov2024commentsintegrabilitysymmetricorbifold}, thermalization \cite{Hampton:2019csz,Hampton:2019hya}, various aspects of black hole microstates\cite{Guo:2021ybz,Guo:2021gqd,Guo:2022and} and scrambling \cite{Chen:2024oqv}, etc.

\subsection{Monodromy}\label{monodromy sub}

Correlation functions involving twist operators exhibit effects arising from monodromy, a phenomenon that occurs when a field is transported around twist operators in a closed loop. These effects were initially developed for $\mathbb Z_N$ orbifolds in \cite{Dixon:1986qv}. Here we provide a brief review of monodromy within the context of our setup.

Let us consider the antisymmetric combination of the two field copies, defined away from the twist operator's location as 
\bea\label{antisymmetric X}
X \equiv {1\over\sqrt2}(X^{(1)} -  X^{(2)})
\eea
This field $X$ is antisymmetric under the interchange of the two copies. 
Now, consider the $z$-plane with insertions of the twist operator $\sigma$.
When $X$ traces a loop around any twist operator, the boundary conditions (\ref{twist bc}) cause $X^{(1)}$ and $X^{(2)}$ to exchange. Consequently, $X$ obtains a minus sign upon a $2\pi$ rotation around a twist operator. Therefore, $X$ is periodic only under a $4\pi$ rotation around the twist operator.

The twist operators create branch cuts of order 2 on the $z$-plane, with $X^{(1)}$ and $X^{(2)}$ representing fields defined on the two distinct Riemann sheets. The field $X$ becomes periodic when encircling a closed loop on the Riemann surface. More generally, $X$ satisfies a periodicity condition along $C$ on the Riemann surface associated with the branched $z$-plane
\cite{Dixon:1986qv}
\bea\label{monodromy}
0=\Delta_{C} X= \oint_{C}dz\partial X + \oint_{C}d\bar z\bar{\partial} X
\eea 
These conditions impose nontrivial constraints that can couple the left and right moving sectors. Such constraints arise from loops on the covering surface that cannot be contracted to a point. In the case of two twist operators the covering surface is a sphere which has no nontrivial loops. Therefore, there are no nontrivial monodromy constraints in this case. The first instance of nontrivial monodromy effects occur with four twist operators, where the covering surface becomes a torus. In this paper, we focus on this scenario, particularly the coupling between left and right movers, which was recently first explored in \cite{Guo:2024edj}.

\subsection{The effect of four twist operators} \label{bogoliubov ansatz}

In this section, we describe various effects produced by four twist operators. In this paper, we focus on the simplest case where only two copies of the seed CFT are present, such that $N=2$. The ground state in the untwisted sector is given by 
\bea
|0\rangle \equiv |0\rangle^{(1)}|0\rangle^{(2)}|\bar 0\rangle^{(1)}|\bar 0\rangle^{(2)}
\eea
where $|0\rangle^{(i)}|\bar 0\rangle^{(i)}$ denotes the vacuum state for each copy $i$. 
Consider an initial state in the untwisted sector containing some left and right moving excitations
\bea \label{initial state}
\alpha_{-n_1}\alpha_{-n_2}\ldots \bar\alpha_{-m_1}\bar\alpha_{-m_2}\ldots|0\rangle 
\eea
where $n_k,m_k>0$. The excitations here are antisymmetric combinations of modes between the two copies
\begin{align}\label{antisymm modes}
\alpha_{p} \equiv {1\over\sqrt2}(\alpha^{(1)}_{p} - \alpha^{(2)}_{p})  \qquad
\bar \alpha_{p} \equiv {1\over\sqrt2}(\bar \alpha^{(1)}_{p} - \bar \alpha^{(2)}_{p})
\end{align}
The symmetric combinations of the modes remain invariant under the orbifolding procedure and thus are not affected by the action of the twist operators: they simply pass through the twist operators \cite{Guo:2022sos,Avery:2010er}.

Let us consider the effect of applying four twist operators on this initial state. The first twist operator joins the two untwisted copies into a twist-2 copy, while the second twist operator splits it back into two untwisted copies; similarly, the third and fourth twist operators repeat this process. After applying all four twist operators, we return to a state in the untwisted sector. We define this final state as $\phi$
\be
|\phi\rangle = \prod_{i=1}^4\sigma(w_i,\bar w_i) \alpha^{(i_1)}_{-n_1}\alpha^{(i_2)}_{-n_2}\ldots \bar\alpha^{(j_1)}_{-m_1}\alpha^{(j_2)}_{-m_2}\ldots|0\rangle
\ee
To determine the state $\phi$, we use the covering map. On the covering surface, the fields are free and follow Wick contraction rules, leading to three basic rules for computing the state
$\phi$:

\b

(i) Contraction: Two modes in the initial state can `contract', producing a constant. Four such contraction terms arise from different combinations of left and right moving modes, defined as follows
\be
C^{\alpha\alpha}[m,n] \qquad   C^{\alpha\bar\alpha}[m,n]  \qquad C^{\bar\alpha\bar\alpha}[m,n]
\ee
For instance, a contraction between two left moving modes $\alpha_{-m}\alpha_{-n}$ produces $C^{\alpha\alpha}[m,n]$, while a contraction between a left moving mode and a right moving mode $\alpha_{-m}\bar\alpha_{-n}$ yields $C^{\alpha\bar\alpha}[m,n]$. In the contraction process, all possible pairs of modes are considered. Each mode either contracts with another mode, producing the constants above, or passes through the twist, as described in step (ii) below.

\b

(ii) Propagation: Any modes that remain after the contraction pass through the twist operator and become new modes under the action of the twist operator, as follows
\begin{align}\label{f}
 \a_{-n} \rightarrow & \sum_{p>0}f^{\alpha\alpha}_{n,p}\alpha_{-p} + \sum_{p>0}f^{\alpha\bar\alpha}_{n,p} \bar{\alpha}_{-p}\nn
\bar\a_{-n}\rightarrow & \sum_{p>0}f^{\bar\alpha\alpha}_{n,p}\alpha_{-p} + \sum_{p>0}f^{\bar\alpha\bar\alpha}_{n,p}\bar\alpha_{-p}
\end{align}

(iii) Pair creation: After the previous two steps, the modes in the initial state have either been contracted or propagated through the twist. What remains is the twist operator acting on the untwisted vacuum, which gives the state
\begin{align}\label{pair}
|\chi\rangle\equiv \, &\prod_{i=1}^4\sigma(w_i,\bar w_i)|0\rangle\nn 
= \, &\mathcal C\exp\Big(\sum_{m,n>0}\gamma_{m,n}(w_i,\bar w_i)\a_{-m}\a_{-n} + \sum_{m,n>0}\beta_{m,n}(w_i,\bar w_i)\alpha_{-m}\bar{\alpha}_{-n}\nn 
&\qquad \quad+\sum_{m,n>0}\bar\gamma_{m,n}(w_i,\bar w_i)\bar\a_{-m}\bar\a_{-n}\Big)|0\rangle
\end{align}
where the coefficient $\mathcal C$ is
\bea 
\mathcal C \equiv\langle 0| \prod_{i=1}^4\sigma(w_i,\bar w_i)|0\rangle
\eea
which has been computed in previous works \cite{Dixon:1986qv,Lunin:2000yv}.
To better understand these rules, consider an example with a single initial mode
\be\label{f_ans_1}
\prod_{i=1}^4\sigma(w_i,\bar w_i)\, \a_{-n}|0\rangle=\big(\sum_{p>0}f^{\alpha\alpha}_{n,p}\alpha_{-p} + \sum_{p>0}f^{\alpha\bar\alpha}_{n,p} \bar{\alpha}_{-p}\big)|\chi\rangle
\ee
We first apply the propagation rule (\ref{f}) to propagate the initial mode through the twist. Then the twist operator acts on the untwisted vacuum to produce pairs according to the pair creation rule (\ref{pair}). Now consider an example with two initial left moving modes 
\begin{align}\label{contr aa}
&\prod_{i=1}^4\sigma(w_i,\bar w_i)\a_{-m}\a_{-n}|0\rangle
\nn
=\, &\bigg(\Big(\sum_{p>0}f^{\alpha\alpha}_{m,p}\alpha_{-p}+\sum_{p>0}f^{\alpha\bar\alpha}_{m,p}\bar\alpha_{-p}\Big)\Big(\sum_{q>0}f^{\alpha\alpha}_{n,q}\alpha_{-q}+\sum_{q>0}f^{\alpha\bar\alpha}_{n,q}\bar\alpha_{-q}\Big) +C^{\alpha\alpha}[m,n]\bigg)|\chi\rangle
\end{align}
The first term in parentheses arises from the propagation of the two initial modes, while the second term comes from the contraction. The state $\chi$, which captures pair creation, is given by (\ref{pair}). Similarly, for an initial state consisting of one left moving mode and one right moving mode 
\begin{align}\label{contr aabar}
&\prod_{i=1}^4\sigma(w_i,\bar w_i)\a_{-m}\bar\a_{-n}|0\rangle
\nn
=\, &\bigg(\Big(\sum_{p>0}f^{ \alpha\alpha}_{m,p}\alpha_{-p}+\sum_{p>0}f^{ \alpha\bar\alpha}_{m,p}\bar\alpha_{-p}\Big)\Big(\sum_{q>0}f^{\bar \alpha\alpha}_{n,q}\alpha_{-q}+\sum_{q>0}f^{\bar \alpha\bar\alpha}_{n,q}\bar\alpha_{-q}\Big) +C^{\alpha\bar\alpha}[m,n]\bigg)|\chi\rangle
\end{align}

In the appendix of \cite{Guo:2024edj}, it was shown that the above rule arises from Wick contraction on the covering space.
We also note that the coefficients $\gamma$ and $\beta$, which characterize pair creation processes, were computed explicitly in \cite{Guo:2024edj}. 
These results are summarized in section \ref{results}, where we have collected all of our results together. Furthermore as we will show, the coefficients of the contractions can be seen morally as the conjugates of the pair creation coefficients.
In this work, we will compute both the propagation and the contraction, thereby completing the full set of effects produced by four twist operators. 

\subsection{Propagation}

Using (\ref{f_ans_1}), which contains a single left moving mode in the initial state, $\alpha_{-n}$, we can apply a left moving mode in the final state, $\alpha_{p}$, which yields
\begin{align}\label{faa}
f^{\alpha\alpha}_{n,p}= {1\over p}{\langle 0| \a_{p} \prod_{i=1}^4\sigma(w_i,\bar w_i)\, \a_{-n} |0\rangle\over \langle0| \prod_{i=1}^4\sigma(w_i,\bar w_i)| 0\rangle}
= {1\over p}{\langle 0| \a_{p} \prod_{i=1}^4\sigma(z_i,\bar z_i)\, \a_{-n} |0\rangle\over \langle0| \prod_{i=1}^4\sigma(z_i,\bar z_i)| 0\rangle}
\end{align}
or a right moving mode $\bar\alpha_{p}$, which yields
\begin{align}\label{f}
f^{\alpha\bar \alpha}_{n,p}&= {1\over p}{\langle 0| \bar \a_{p} \prod_{i=1}^4\sigma(w_i,\bar w_i)\, \a_{-n} |0\rangle\over \langle0| \prod_{i=1}^4\sigma(w_i,\bar w_i)| 0\rangle}
= {1\over p}{\langle 0| \bar \a_{p} \prod_{i=1}^4\sigma(z_i,\bar z_i)\, \a_{-n} |0\rangle\over \langle0| \prod_{i=1}^4\sigma(z_i,\bar z_i)| 0\rangle}
\end{align} 
where we have mapped the amplitudes to the $z$-plane using $z=e^{w}$. Note that the Jacobian factors arising from this mapping of the twist operators cancel out between the numerator and denominator.
The mode before the twist, $\a_{-n}$, is defined by the contour at $\tau<\tau_i$, corresponding to $|z|<|z_i|$, while the modes after the twist, $\a_{p}$ and $\bar \a_{p}$, are defined by the contour at $\tau>\tau_i$, corresponding to $|z|>|z_i|$. 
By inserting the modes (\ref{alpha exp}), (\ref{antisymm modes}) mapped to the appropriate $z$-plane locations, into the above correlation functions (\ref{faa}) and (\ref{f}), we obtain
\begin{subequations}
\begin{align}
\label{faa z plane}
f^{\alpha\alpha}_{n,p}&=-{1\over p}\oint_{|z|>|z_i|}{dz\over2\pi}z^p\oint_{|z'|<|z_i|}{dz'\over2\pi }z'^{-n}g(z,z';z_i,\bar z_i)
\\
\label{pair creation z plane aabar}
f^{\alpha\bar \alpha}_{n,p}&=-{1\over p}\oint_{|\bar z|>|\bar z_i|}{d\bar z\over2\pi }\bar z^p\int_{|z'|<|z_i|}{d z'\over2\pi }z'^{-n}b(\bar z, z';z_i,\bar z_i)
\end{align}
\end{subequations}
where the correlation functions $g$ and $b$ are defined as
\begin{subequations}
\begin{align}
\label{g}
g(z,z';z_i,\bar z_i) 
&\equiv-{\langle 0|\partial_z X\partial_{z'} X\prod_{i=1}^4\sigma(z_i,\bar z_i) |0\rangle\over \langle 0| \prod_{i=1}^4\sigma(z_i,\bar z_i)|0\rangle}\\
\label{b}
b(\bar z, z';z_i,\bar z_i) 
&\equiv-{\langle 0|\partial_{\bar z} X\partial_{ z'} X\prod_{i=1}^4\sigma(z_i,\bar z_i) |0\rangle\over \langle 0| \prod_{i=1}^4\sigma(z_i,\bar z_i)|0\rangle}
\end{align}
\end{subequations}
We note that $b$ measures correlations between left and right moving fields, an effect that doesn't arise in the two twist case.
The expressions for $f^{\bar\alpha\alpha}_{n,p}$ and $f^{\bar\alpha\bar\alpha}_{n,p}$ can be obtained by simply switching the barred with unbarred coordinates in the expressions above.

\subsection{Contraction}

To derive the analogous expressions for the contractions, we can act on the states (\ref{contr aa}) and (\ref{contr aabar}) with just the vacuum $\langle 0|$ which gives the following expressions
\begin{align}\label{Caa Caabar z}
C^{\alpha\alpha}[m,n]&= {\langle 0|  \prod_{i=1}^4\sigma(w_i,\bar w_i)\a_{-m}\a_{-n} |0\rangle\over \langle0| \prod_{i=1}^4\sigma(w_i,\bar w_i)| 0\rangle}
= {\langle 0|  \prod_{i=1}^4\sigma(z_i,\bar z_i)\a_{-m}\a_{-n} |0\rangle\over \langle0| \prod_{i=1}^4\sigma(z_i,\bar z_i)| 0\rangle}
\nn
C^{\alpha\bar\alpha}[m,n]&= {\langle 0|  \prod_{i=1}^4\sigma(w_i,\bar w_i)\a_{-m}\bar \a_{-n} |0\rangle\over \langle0| \prod_{i=1}^4\sigma(w_i,\bar w_i)| 0\rangle}
= {\langle 0|  \prod_{i=1}^4\sigma(z_i,\bar z_i)\a_{-m}\bar \a_{-n} |0\rangle\over \langle0| \prod_{i=1}^4\sigma(z_i,\bar z_i)| 0\rangle}
\end{align}
where we have mapped them to the $z$-plane using the transformation $z=e^{w}$. The modes are placed before the twist operators, defined by a contour at $\tau<\tau_i$, which corresponds to $|z|<|z_i|$.
Again substituting the mode definitions from (\ref{alpha exp}) and (\ref{antisymm modes}) at the appropriate $z$-plane locations into the above expressions yields the contractions
\begin{align}\label{Caa Caabar}
C^{\alpha\alpha}[m,n]&=-\oint_{|z|<|z_i|}{dz\over 2\pi}z^{-m}\oint_{|z'|<|z_i|}{dz'\over 2\pi}z'^{-n}g(z,z';z_i,\bar z_i)
\nn
\nn
C^{\alpha\bar\alpha}[m,n]&=-\oint_{|z'|<|z_i|}{dz'\over 2\pi}z'^{-m}\oint_{|\bar z|<|\bar z_i|}{d\bar z\over 2\pi}\bar z^{-n}b(\bar z , z'; z_1, \bar z_2)
\end{align}
where $C^{\bar\alpha\bar\alpha}[m,n]$ is simply the barred version of $C^{\alpha\alpha}[m,n]$ where one can simply replace the barred quantities with the unbarred quantities. 

\subsection{Review of the correlators}\label{summary}

The effects discussed above are governed by two correlation functions: $g(z,z';z_i,\bar z_i)$ and $b(\bar z,z';z_i,\bar z_i)$. The general expressions for these correlators at arbitrary twist locations $z_i,\bar z_i$ are provided in \cite{Guo:2024edj}, building upon the specific cases discussed in \cite{Dixon:1986qv}. In this section, we summarize the results and outline the method used to derive them, while referring the reader to \cite{Guo:2024edj} for further technical details.

The correlators $g$ and $b$ are given by
\begin{align}\label{g and b z plane}
&g(z,z';z_i,\bar z_i)\nn 
=\ &{1\over2}{(z-z_1)^{1\over2}(z-z_3)^{1\over2}(z'-z_2)^{1\over2}(z'-z_4)^{1\over2}\over(z-z_2)^{1\over2}(z-z_4)^{1\over2}(z'-z_1)^{1\over2}(z'-z_3)^{1\over2}}{1\over(z-z')^2}
\nn
&+ {1\over2}{(z-z_2)^{1\over2}(z-z_4)^{1\over2}(z'-z_1)^{1\over2}(z'-z_3)^{1\over2}\over(z-z_1)^{1\over2}(z-z_3)^{1\over2}(z'-z_2)^{1\over2}(z'-z_4)^{1\over2}}{1\over(z-z')^2}
\nn
&+ {A(z_i,\bar z_i)\over  (z-z_1)^{1\over2} (z-z_2)^{1\over2} (z-z_3)^{1\over2} (z-z_4)^{1\over2} (z'-z_1)^{1\over2} (z'-z_2)^{1\over2} (z'-z_3)^{1\over2} (z'-z_4)^{1\over2}}
\nn 
\nn
&b (\bar z,z'; z_i,\bar z_i)\nn 
=\ &{B(z_i,\bar z_i)\over  (\bar z-\bar z_1)^{1\over2} (\bar z-\bar z_2)^{1\over2} (\bar z-\bar z_3)^{1\over2} (\bar z-\bar z_4)^{1\over2} (z'-z_1)^{1\over2} (z'-z_2)^{1\over2} (z'-z_3)^{1\over2} (z'-z_4)^{1\over2}}
\end{align}
The functions $A(z_i,\bar z_i)$ and $B(z_i,\bar z_i)$ are defined as
\begin{align}\label{ABz}
A(z_i,\bar z_i)&\equiv -A(x,\bar x)(z_1-z_3)(z_2-z_4)
\nn 
B(z_i,\bar z_i)&\equiv B(x,\bar x)|z_1-z_3||z_2-z_4|
\end{align}
where the cross ratio $x$ is given by
\begin{equation}
    x = {(z_1-z_2)(z_3-z_4)\over(z_1-z_3)(z_2-z_4)}
\end{equation}
The functions $A(x,\bar x)$ and $B(x,\bar x)$ are expressed as
\begin{align} \label{ABx}
A(x,\bar x) &= x(1-x){d\over dx}\ln(F(x)\bar F(1-\bar x)+F(1-x)\bar F(\bar x))
\nn 
 B(x,\bar x) &={1\over\pi (F(x)\bar F(1-\bar x)+F(1-x)\bar F(\bar x))} 
\end{align}
 where $F(x)$ is the hypergeometric function
 \begin{align}
F(x)\equiv {}_2F_1\Big({1\over2},{1\over2};1;x\Big)
 \end{align}
The functions $A(x,\bar x)$ and $B(x,\bar x)$ can also be written as
\begin{align} \label{AB}
A(x,\bar x) &= {(E(x) - (1-x)K(x))\bar K(1-\bar x)-( E(1-x) - xK(1-x) )\bar K(\bar x) \over2 (K(x)\bar K(1-\bar x)+  K(1-x)\bar K(\bar x))}
\nn 
 B(x,\bar x) &= {\pi\over 4(K(x)\bar K(1-\bar x)+  K(1-x)\bar K(\bar x))}
\end{align}
where $K(y)$ and $E(y)$ denote the complete elliptic integrals of the first and second kinds. This form will be used for explicit computations in later sections.

In the following, we outline the steps to compute the correlation functions $g(z,z';z_i,\bar z_i)$ and $b(\bar z,z';z_i,\bar z_i)$.

\paragraph{1. Map to the covering space.} First consider the twist operator locations at the special positions $z_1=0, z_2= x, z_3= 1, z_4 =\infty$. In this setup, the covering map can be found explicitly and is given by
\bea \label{t plane}
z = {\mathcal P(t) - e_1\over e_2 -e_1}
\eea
where $\mathcal P$ is the Weierstrass $\mathcal P$ function, and $e_i$'s are its values at the half-periods of the torus. This maps the base space, which is a sphere, into a torus. The torus modulus parameter $\tau$ is then related to the cross ratio $x$.

\paragraph{2. Two point function on the torus.} Mapping to the covering space resolves the branch points associated with the twist operators. As a result, the correlation functions $g(z,z';z_i,\bar z_i)$ and $b(\bar z,z';z_i,\bar z_i)$, which involve two free fields and four twist operators, transform into two-point functions of free fields on the torus. This two point function is well-known in the literature. 

To reproduce the double pole when the two points are close on the torus, the two-point functions are uniquely determined by the Weierstrass $\mathcal P$ function up to two additional constants. These constants are fixed by imposing the condition that the fields are periodic on the torus along both periods $(1,\tau)$. On the base space, these periodicity conditions correspond to the monodromy conditions (\ref{monodromy}).

\paragraph{3, Correlation function on the base space.} Thanks to the form of the covering map, the two point function, which depends on the torus coordinate $t$ and the torus modulus parameter $\tau$, can be expressed in terms of the original base space coordinate $z$ and the cross ratio $x$. 

\paragraph{4. Correlation function at arbitrary twist locations.} Once the correlation functions at specific twist locations are obtained, an SL$(2,\mathbb C)$ transformation can be applied to move the twist operators to arbitrary locations $z_i$. This yields the correlation functions $g(z,z';z_i,\bar z_i)$ and $b(\bar z,z';z_i,\bar z_i)$.

\subsection{Properties of the correlators}\label{sec pc}

There are two important properties of the correlation functions $g(z,z';z_i,\bar z_i)$ and $b(\bar z,z';z_i,\bar z_i)$ that will be useful. For further technical details, see \cite{Guo:2024edj}.

\paragraph{Twist interchange symmetries}

Since the four twist operators are identical, the correlation functions $g(z,z';z_i,\bar z_i)$ and $b(\bar z,z';z_i,\bar z_i)$ are invariant under the interchange of twist locations, i.e., $z_i\leftrightarrow z_j$. For an explicit verification of this symmetry, see section 3.5 in \cite{Guo:2024edj}. Consequently, the contraction, propagation and pair creation coefficients are completely symmetric among the four twist operator locations. 

\paragraph{Reduction to two twist operators}

The correlation functions $g(z,z';z_i,\bar z_i)$ and $b(\bar z,z';z_i,\bar z_i)$, which involve four twist operators, can be reduced to a correlation function with two twist operators when any two of the four twist operators are brought together. For example, when $z_3 \to z_4$, the term $\prod_{i=1}^4\sigma(z_i,\bar z_i)$ in (\ref{g}) and (\ref{b}) is replaced by $\sigma(z_2,\bar z_2)\sigma(z_1,\bar z_1)$. For an explicit verification of this reduction, see section 3.6 in \cite{Guo:2024edj}. As a result, the contraction, propagation and pair creation coefficients for four twist operators reduce to their corresponding values for two twist operators. 

\section{Propagation for four twists}\label{twist effects}

In this section, we derive the propagation coefficients for four twist operators, using the correlators reviewed in the previous section. 

\subsection{Left moving propagation}
 
We begin by recalling the left moving propagation coefficients given in (\ref{faa z plane}) 
\bea 
f^{\alpha\alpha}_{n,p}=-{1\over p} \oint_{|z|>|z_i|}{dz\over 2\pi }z^p\oint_{|z'|<|z_i|}{dz'\over2\pi }z'^{-n}g(z,z';z_i,\bar z_i)
\eea
Inserting the expression for the amplitude $g$ from (\ref{g and b z plane}) and separating it into three terms, we obtain  
\begin{align}\label{I II III}
f^{\alpha\alpha}_{n,p}
&\equiv f^{\alpha\alpha}_{\text{I}} + f^{\alpha\alpha}_{\text{II}} + f^{\alpha\alpha}_{\text{III}}
\end{align}
where the three terms are defined as follows
\begin{align}\label{fI fII fIII}
f^{\alpha\alpha}_{\text I}&=  -{1\over 2p}\oint_{|z|>|z_i|}{dz\over2\pi } z^p\oint_{|z'|<|z_i|}{dz'\over 2\pi }  z'^{-n} ~{(z-z_1)^{1\over2}(z-z_3)^{1\over2}(z'-z_2)^{1\over2}(z'-z_4)^{1\over2}\over(z-z_2)^{1\over2}(z-z_4)^{1\over2}(z'-z_1)^{1\over2}(z'-z_3)^{1\over2}}{1\over(z-z')^2}\nn
\nn
f^{\alpha\alpha}_{\text{II}}&=- {1\over 2p}\oint_{|z|>|z_i|}{dz\over2\pi } z^p\oint_{|z'|<|z_i|}{dz'\over2\pi }  z'^{-n}{(z-z_2)^{1\over2}(z-z_4)^{1\over2}(z'-z_1)^{1\over2}(z'-z_3)^{1\over2}\over(z-z_1)^{1\over2}(z-z_3)^{1\over2}(z'-z_2)^{1\over2}(z'-z_4)^{1\over2}}{1\over(z-z')^2}\nn
\nn
f^{\alpha\alpha}_{\text{III}}&= {1\over p}A(x,\bar x)(z_1-z_3)(z_2-z_4)\oint_{|z|>|z_i|}{dz\over2\pi } z^p\oint_{|z'|<|z_i|}{dz'\over2\pi }  z'^{-n}\nn 
&\quad\times{1\over  (z-z_1)^{1\over2} (z-z_2)^{1\over2} (z-z_3)^{1\over2} (z-z_4)^{1\over2} (z'-z_1)^{1\over2} (z'-z_2)^{1\over2} (z'-z_3)^{1\over2} (z'-z_4)^{1\over2}}
\end{align}
Note that the first and second terms are related by a simple interchange of the twist locations
\bea\label{I II switch}
f^{\alpha\alpha}_{\text{II}} = f^{\alpha\alpha}_{\text{I}}(z_1\leftrightarrow z_2,z_3\leftrightarrow z_4)
\eea
In the forthcoming computations, we evaluate the contours for $|z|>|z_i|$ at $z=\infty$ and for $|z|<|z_i|$ at $z=0$, as these contours can be freely deformed to these respective locations.
For $z\to\infty$, where $|z|\gg|z_i|$, we expand the integrand factors as
\bea\label{expansion}
(z-z_i)^{\pm{1\over2}}=z^{\pm{1\over2}}(1-z_{i}z^{-1})^{\pm{1\over2}}=\sum_{k\geq0}{}^{\pm{1\over2}}C_{k}(-1)^kz^k_{i}z^{-k\pm{1\over2}}
\eea
where ${}^{p}C_q$ is the binomial coefficient
\bea
{}^{p}C_q={p!\over q! (p-q)!}
\eea
Similarly, for $z\to 0$, where $|z|\ll|z_i|$, we expand the integrand as
\bea\label{expansion 0}
(z-z_i)^{\pm{1\over2}}=(-z_i)^{\pm{1\over2}}(1-z_i^{-1}z)^{\pm{1\over2}}=(-1)^{\pm{1\over2}}\sum_{k\geq0}{}^{\pm{1\over2}}C_{k}(-1)^kz^{-k\pm{1\over2}}_iz^k
\eea
Utilizing these expansions, next we compute each term in the propagation.

\paragraph{Term I and II}

To compute the first term in the propagation $f^{\alpha\alpha}_{\text I}$, we substitute the expansions (\ref{expansion}) and (\ref{expansion 0}) into (\ref{fI fII fIII})
\begin{align}\label{f I}
f^{\alpha\alpha}_{\text{I}} =& -{1\over 2p}\bigg({z_2z_4\over z_1z_3}\bigg)^{1\over2}\sum_{k_i,k'_i \geq0}{}^{{1\over2}}C_{k_1}{}^{-{1\over2}}C_{k_2}{}^{{1\over2}}C_{k_3}{}^{-{1\over2}}C_{k_4}{}^{-{1\over2}}C_{k'_1}{}^{{1\over2}}C_{k'_2}{}^{-{1\over2}}C_{k'_3}{}^{{1\over2}}C_{k'_4}\nn 
&(-1)^{\sum k_i+\sum k'_i}z^{k_1-k'_1}_{1}z^{k_2-k'_2}_{2}z^{k_3-k'_3}_{3}z^{k_4-k'_4}_{4}\oint_{z=\infty}{dz\over 2\pi } z^{p-\sum k_i}\oint_{z'=0}{dz'\over2\pi }   z'^{\sum k'_i-n}~{1\over(z-z')^2}
\end{align}
Focusing on the integrations in the last line, we first perform the contour integral around the point $z'=0$, which yields
\begin{align}
(n-\sum k'_i)i\oint_{z=\infty}{dz\over 2\pi } z^{p-n-\sum k_i+\sum k'_i-1}\bigg|_{n-\sum k'_i>0}
\end{align}
To compute the contour integral around $z=\infty$, we make the coordinate transformation
$z={1\over u}$, which maps the pole at $z=\infty$ to a pole at $u=0$. This yields the expression
\begin{align}
(n-\sum k'_i)i\oint_{u=0}{du\over 2\pi }  u^{-p+n+\sum k_i-\sum k'_i-1}\bigg|_{n-\sum k'_i>0}
= -(n-\sum k'_i) \d_{p-n-\sum k_i+\sum k'_i,0}\bigg|_{n-\sum k'_i>0}
\end{align}
Substituting this result back, we find
\begin{align}
f^{\alpha\alpha}_{\text{I}} =& {1\over 2p}\bigg({z_2z_4\over z_1z_3}\bigg)^{1\over2}\sum_{k_i,k'_i\in D}
{}^{{1\over2}}C_{k_1}{}^{-{1\over2}}C_{k_2}{}^{{1\over2}}C_{k_3}{}^{-{1\over2}}C_{k_4}{}^{-{1\over2}}C_{k'_1}{}^{{1\over2}}C_{k'_2}{}^{-{1\over2}}C_{k'_3}{}^{{1\over2}}C_{k'_4}\nn 
&(-1)^{\sum k_i+\sum k'_i}z^{k_1-k'_1}_{1}z^{k_2-k'_2}_{2}z^{k_3-k'_3}_{3}z^{k_4-k'_4}_{4}(n-\sum k'_i)
\end{align}
where the sum is taken over the region satisfying
\bea
D = \{k_i,k'_i\geq 0\ ,\quad n-\sum k'_i>0\ ,\quad p-n-\sum k_i+\sum k'_i=0\}
\eea
From the third condition above, we choose 
\begin{equation}
k_4=p-n-(k_1+k_2+k_3)+k'_1+k'_2+k'_3+k'_4
\end{equation}
Imposing the first and second conditions, we then deduce the following constraints on the remaining indices
\begin{align}
k_3&\leq p-n-k_1-k_2+k'_1+k'_2+k'_3+k'_4 \nn
k_2&\leq p-n-k_1+k'_1+k'_2+k'_3+k'_4 \nn
k_1&\leq p-n+k'_1+k'_2+k'_3+k'_4 \nn
k'_4&\leq n-(k'_1+k'_2+k'_3+1)\nn
k'_3&\leq n-(k'_1+k'_2+1)\nn 
k'_2&\leq n-(k'_1+1)\nn 
k'_1&\leq n-1
\end{align}
with the additional constraint that
\begin{equation}
k'_4\geq n-p-(k'_1+k'_2+k'_3)
\end{equation}
which yields
\begin{align}\label{fI}
f^{\alpha\alpha}_{\text{I}} =\, & {1\over 2p}\bigg({z_2z_4\over z_1z_3}\bigg)^{1\over2}(-1)^{p+n}\sum_{k'_1=0}^{n-1}\sum_{k'_2=0}^{n-k'_1-1}\sum_{k'_3=0}^{n-k'_1-k'_2-1}\sum_{k'_4=\max[0,n-p-k'_1-k'_2-k'_3]}^{n-k'_1-k'_2-k'_3-1}\nn 
&\sum_{k_1=0}^{p-n+k'_1+k'_2+k'_3+k'_4}~\sum_{k_2=0}^{p-n-k_1+k'_1+k'_2+k'_3+k'_4}~\sum_{k_3=0}^{p-n-k_1-k_2+k'_1+k'_2+k'_3+k'_4}(n-k'_1-k'_2-k'_3-k'_4)\nn 
&{}^{-{1\over2}}C_{k'_1}{}^{{1\over2}}C_{k'_2}{}^{-{1\over2}}C_{k'_3}{}^{{1\over2}}C_{k'_4}{}^{{1\over2}}C_{k_1}{}^{-{1\over2}}C_{k_2}{}^{{1\over2}}C_{k_3}{}^{-{1\over2}}C_{p-n-k_1-k_2-k_3+k'_1+k'_2+k'_3+k'_4}\nn 
&z^{k_1-k'_1}_{1}z^{k_2-k'_2}_{2}z^{k_3-k'_3}_{3}z^{p-n-k_1-k_2-k_3+k'_1+k'_2+k'_3}_{4}
\end{align}

For the second term of propagation, $f^{\alpha\alpha}_{\text{II}}$, we can simply make the switches $z_1\leftrightarrow z_2,z_3\leftrightarrow z_4$ as indicated in (\ref{I II switch}).

\paragraph{Term III}

Let's now compute the third term of propagation, $f^{\alpha\alpha}_{\text{III}}$.  
Inserting the expansions (\ref{expansion}) and (\ref{expansion 0}) into the third term in (\ref{fI fII fIII}), we obtain
\begin{align}\label{Term 3}
f^{\alpha\alpha}_{\text{III}}=\, &{A(x,\bar x)(z_1-z_3)(z_2-z_4)\over \sqrt{z_1z_2z_3z_4}}\nn 
&\times\Big({1\over p}\sum_{k_i\geq0}{}^{-{1\over2}}C_{k_1}{}^{-{1\over2}}C_{k_2}{}^{-{1\over2}}C_{k_3}{}^{-{1\over2}}C_{k_4}(-1)^{\sum k_i}z_1^{k_1}z_2^{k_2}z_3^{k_3}z_4^{k_4}\oint_{z=\infty}{dz\over2\pi } z^{p-\sum k_i-2}\Big)\nn
&\times\Big(\sum_{k'_i\geq0}{}^{-{1\over2}}C_{k'_1}{}^{-{1\over2}}C_{k'_2}{}^{-{1\over2}}C_{k'_3}{}^{-{1\over2}}C_{k'_4}(-1)^{\sum k'_i}z_1^{-k'_1}z_2^{-k'_2}z_3^{-k'_3}z_4^{-k'_4}\oint_{z'=0}{dz'\over2\pi }  z'^{\sum k'_i-n}\Big) 
\end{align}
where the contour integrals can be evaluated as follows
\begin{align}
\oint_{z=\infty}{dz\over2\pi } z^{p-\sum k_i-2} =\, & \oint_{u=0}{du\over2\pi } u^{-p+\sum k_i}
=i\d_{p-\sum k_i-1,0}\nn
\oint_{z'=0}{dz'\over2\pi }  z'^{\sum k'_i-n} =\, & i\d_{n-\sum k'_i-1,0}
\end{align}
This gives us
\begin{align}
f^{\alpha\alpha}_{\text{III}}=\, &-{A(x,\bar x)(z_1-z_3)(z_2-z_4)\over \sqrt{z_1z_2z_3z_4}}
\nn 
&\quad\times\bigg(\ {1\over p}\sum_{\substack{k_i\geq0\\p-\sum k_i-1=0}}{}^{-{1\over2}}C_{k_1}{}^{-{1\over2}}C_{k_2}{}^{-{1\over2}}C_{k_3}{}^{-{1\over2}}C_{k_4}(-1)^{\sum k_i}z_1^{k_1}z_2^{k_2}z_3^{k_3}z_4^{k_4}
 \bigg)\nn
&\quad\times\bigg(\sum_{\substack{k'_i\geq0\\n-\sum k'_i-1=0}}{}^{-{1\over2}}C_{k'_1}{}^{-{1\over2}}C_{k'_2}{}^{-{1\over2}}C_{k'_3}{}^{-{1\over2}}C_{k'_4}(-1)^{\sum k'_i}z_1^{-k'_1}z_2^{-k'_2}z_3^{-k'_3}z_4^{-k'_4} \bigg) 
\end{align}
The summation region can be rewritten as
\begin{align}
p-k_1-k_2-k_3-k_4-1&=0\implies k_4=p-k_1-k_2-k_3-1\nn 
k_4&\geq 0 \implies p-k_1-k_2-1\geq k_3\nn
k_3&\geq 0 \implies p-k_1-1\geq k_2\nn
k_2&\geq 0 \implies p-1\geq k_1\nn
\nn
n-k'_1-k'_2-k'_3-k'_4-1&=0\implies k'_4=n-k'_1-k'_2-k'_3-1
\nn
k'_4&\geq 0 \implies n-k'_1-k'_2-1\geq k'_3\nn
k'_3&\geq 0 \implies n-k'_1-1\geq k'_2\nn
k'_2&\geq 0 \implies n-1\geq k'_1
\end{align}
which gives
\begin{align}\label{fIII}
f^{\alpha\alpha}_{\text{III}}=\, &-{A(x,\bar x)(z_1-z_3)(z_2-z_4)\over \sqrt{z_1z_2z_3z_4}}\nn 
&\quad\times\bigg({(-1)^p\over p}\sum_{k_1=0}^{p-1}\sum_{k_2=0}^{p-k_1-1}\sum_{k_3=0}^{p-k_1-k_2-1}{}^{-{1\over2}}C_{k_1}{}^{-{1\over2}}C_{k_2}{}^{-{1\over2}}C_{k_3}{}^{-{1\over2}}C_{p-k_1-k_2-k_3-1}\nn
&\quad\qquad z_1^{k_1}z_2^{k_2}z_3^{k_3}z_4^{p-k_1-k_2-k_3-1}\bigg)\nn
&\quad\times\bigg((-1)^n\sum_{k'_1=0}^{n-1}\sum_{k'_2=0}^{n-k'_1-1}\sum_{k'_3=0}^{n-k'_1-k'_2-1}{}^{-{1\over2}}C_{k'_1}{}^{-{1\over2}}C_{k'_2}{}^{-{1\over2}}C_{k'_3}{}^{-{1\over2}}C_{n-k'_1-k'_2-k'_3-1}
\nn
&\quad\qquad z_1^{-k'_1}z_2^{-k'_2}z_3^{-k'_3}z_4^{-n+k'_1+k'_2+k'_3+1} \bigg) 
\end{align}

\paragraph{Final expression for the left moving propagation}

Inserting the three terms (\ref{fI}), (\ref{I II switch}) and (\ref{fIII}) into (\ref{I II III}), we obtain
\begin{align}\label{f aa}
f^{\alpha\alpha}_{n,p} =\, & {1\over 2p}(-1)^{p+n}\sum_{k'_1=0}^{n-1}\sum_{k'_2=0}^{n-k'_1-1}\sum_{k'_3=0}^{n-k'_1-k'_2-1}\sum_{k'_4=\max[0,n-p-k'_1-k'_2-k'_3]}^{n-k'_1-k'_2-k'_3-1}\nn 
&\sum_{k_1=0}^{p-n+k'_1+k'_2+k'_3+k'_4}~\sum_{k_2=0}^{p-n-k_1+k'_1+k'_2+k'_3+k'_4}~\sum_{k_3=0}^{p-n-k_1-k_2+k'_1+k'_2+k'_3+k'_4}(n-k'_1-k'_2-k'_3-k'_4)\nn 
&{}^{-{1\over2}}C_{k'_1}{}^{{1\over2}}C_{k'_2}{}^{-{1\over2}}C_{k'_3}{}^{{1\over2}}C_{k'_4}{}^{{1\over2}}C_{k_1}{}^{-{1\over2}}C_{k_2}{}^{{1\over2}}C_{k_3}{}^{-{1\over2}}C_{p-n-k_1-k_2-k_3+k'_1+k'_2+k'_3+k'_4}\nn 
&~~\times\bigg(\Big({z_2z_4\over z_1z_3}\Big)^{1\over2}z^{k_1-k'_1}_{1}z^{k_2-k'_2}_{2}z^{k_3-k'_3}_{3}z^{p-n-k_1-k_2-k_3+k'_1+k'_2+k'_3}_{4}\nn 
&~~~~~~+ \Big({z_1z_3\over z_2z_4}\Big)^{1\over2} z^{k_1-k'_1}_{2}z^{k_2-k'_2}_{1}z^{k_3-k'_3}_{4}z^{p-n-k_1-k_2-k_3+k'_1+k'_2+k'_3}_{3} \bigg)
\nn 
&- {A(x,\bar x)(z_1-z_3)(z_2-z_4)\over \sqrt{z_1z_2z_3z_4}}\nn 
&~~\times\bigg({(-1)^p\over p}\sum_{k_1=0}^{p-1}\sum_{k_2=0}^{p-k_1-1}\sum_{k_3=0}^{p-k_1-k_2-1}{}^{-{1\over2}}C_{k_1}{}^{-{1\over2}}C_{k_2}{}^{-{1\over2}}C_{k_3}{}^{-{1\over2}}C_{p-k_1-k_2-k_3-1}\nn
&\qquad z_1^{k_1}z_2^{k_2}z_3^{k_3}z_4^{p-k_1-k_2-k_3-1}\bigg)\nn
&~~\times\bigg((-1)^n\sum_{k'_1=0}^{n-1}\sum_{k'_2=0}^{n-k'_1-1}\sum_{k'_3=0}^{n-k'_1-k'_2-1}{}^{-{1\over2}}C_{k'_1}{}^{-{1\over2}}C_{k'_2}{}^{-{1\over2}}C_{k'_3}{}^{-{1\over2}}C_{n-k'_1-k'_2-k'_3-1}\nn
&\qquad z_1^{-k'_1}z_2^{-k'_2}z_3^{-k'_3}z_4^{-n+k'_1+k'_2+k'_3+1} \bigg) 
\end{align}
The right moving propagation follows a similar derivation and can be obtained by taking the complex conjugate of the above expression
\be
f^{\bar\alpha\bar\alpha}_{n,p} = (f^{\alpha\alpha}_{n,p})^*
\ee
This means replacing barred quantities with unbarred ones and vice versa, $z_i\leftrightarrow \bar z_i$. The final result is summarized in section \ref{results}.

\subsection{left-right moving propagation}

In this section, we compute the propagation coefficients $f^{\alpha\bar\alpha}_{n,p}$, which describe a left moving excitation being converted into a right moving excitation.
The computation closely parallels the derivation of $f^{\alpha\alpha}_{\text{III}}$ in the previous section. We start with (\ref{pair creation z plane aabar})
\bea 
f^{\alpha\bar\alpha}_{n,p} = -{1\over p}\oint_{\bar z=\infty}{d\bar z\over2\pi }\bar z^{p}\oint_{ z'=0}{dz'\over2\pi }z'^{-n}\, b(\bar z, z';z_i,\bar z_i)
\eea
Inserting the correlation function (\ref{g and b z plane}) yields
\begin{align}
f^{\alpha\bar\alpha}_{n,p} =\,& -{B(z_i,\bar z_i)\over p}\bigg(\oint_{\bar z=\infty}{d\bar z\over2\pi }\bar z^{p}{1\over  (\bar z-\bar z_1)^{1\over2} (\bar z-\bar z_2)^{1\over2} (\bar z-\bar z_3)^{1\over2} (\bar z-\bar z_4)^{1\over2}}\bigg)
\nn
&\qquad \times\bigg(\oint_{ z'=0}{dz'\over2\pi }z'^{-n}
{1\over(z'-z_1)^{1\over2} (z'-z_2)^{1\over2} (z'-z_3)^{1\over2} (z'-z_4)^{1\over2}}\bigg)
\end{align}
The contour integration and the summation index constraints follow in a similar manner as for $f^{\alpha\alpha}_{\text{III}}$, and we thus obtain
\begin{align}\label{f aabar}
f^{\alpha\bar\alpha}_{n,p} =\, &  {B(x,\bar x)|z_1-z_3||z_2-z_4| \over \sqrt{z_1z_2z_3z_4}}\nn 
&\times\bigg({(-1)^p\over p}\sum_{k_1=0}^{p-1}\sum_{k_2=0}^{p-k_1-1}\sum_{k_3=0}^{p-k_1-k_2-1}{}^{-{1\over2}}C_{k_1}{}^{-{1\over2}}C_{k_2}{}^{-{1\over2}}C_{k_3}{}^{-{1\over2}}C_{p-k_1-k_2-k_3-1}\nn
&\qquad\bar z_1^{k_1}\bar z_2^{k_2}\bar z_3^{k_3}\bar z_4^{p-k_1-k_2-k_3-1}\bigg)\nn
&\times\bigg((-1)^n\sum_{k'_1=0}^{n-1}\sum_{k'_2=0}^{n-k'_1-1}\sum_{k'_3=0}^{n-k'_1-k'_2-1}{}^{-{1\over2}}C_{k'_1}{}^{-{1\over2}}C_{k'_2}{}^{-{1\over2}}C_{k'_3}{}^{-{1\over2}}C_{n-k'_1-k'_2-k'_3-1}\nn
&\qquad z_1^{-k'_1}z_2^{-k'_2}z_3^{-k'_3}z_4^{-n+k'_1+k'_2+k'_3+1} \bigg)
\end{align}
The propagation $f^{\bar\alpha\alpha}_{n,p}$ follows a similar derivation and can be obtained by taking the complex conjugate of the above expression
\be
f^{\bar\alpha\alpha}_{n,p} = (f^{\alpha\bar\alpha}_{n,p})^*
\ee
This means that we replace barred quantities with unbarred ones and vice versa, $z_i\leftrightarrow \bar z_i$. The final result is summarized in section \ref{results}.

\section{Contraction for four twists} \label{twist effects contraction}

In this section, we derive the contraction coefficients for four twist operators, using the correlators (\ref{g and b z plane}).

\subsection{Left moving contraction}

We begin by recalling the left moving contraction coefficients as given in (\ref{Caa Caabar})
\begin{align}
C^{\alpha\alpha}[m,n]&=-\oint_{z=0,\, |z|>|z'|}{dz\over 2\pi}z^{-m}\oint_{z'=0}{dz'\over 2\pi}z'^{-n}g(z,z';z_i,\bar z_i)
\end{align}
where the poles encircled by both contours are at $0$, and we have chosen a particular ordering, $|z|>|z'|$, to perform the computation though the result does not depend on this specific choice.
Inserting the expression for the correlation function $g$, recorded in (\ref{g and b z plane}), yields an expression with three terms
\begin{equation}\label{Caa}
    C^{\alpha\alpha}[m,n] = C^{\alpha\alpha}_{\text I} + C^{\alpha\alpha}_{\text{II}} + C^{\alpha\alpha}_{\text{III}}
\end{equation}
which are defined as
\begin{align} \label{CI CII CIII}
C^{\alpha\alpha}_{\text I}
&\equiv-{1\over2}\oint_{z=0,\, |z|>|z'|}{dz\over 2\pi}z^{-m}\oint_{z'=0}{dz'\over 2\pi}z'^{-n}
{(z-z_1)^{1\over2}(z-z_3)^{1\over2}(z'-z_2)^{1\over2}(z'-z_4)^{1\over2}\over(z-z_2)^{1\over2}(z-z_4)^{1\over2}(z'-z_1)^{1\over2}(z'-z_3)^{1\over2}}{1\over(z-z')^2}
\nn
\nn
C^{\alpha\alpha}_{\text{II}}&\equiv-{1\over2}\oint_{z=0,\, |z|>|z'|}{dz\over 2\pi}z^{-m}\oint_{z'=0}{dz'\over 2\pi}z'^{-n}
{(z-z_2)^{1\over2}(z-z_4)^{1\over2}(z'-z_1)^{1\over2}(z'-z_3)^{1\over2}\over(z-z_1)^{1\over2}(z-z_3)^{1\over2}(z'-z_2)^{1\over2}(z'-z_4)^{1\over2}}{1\over(z-z')^2}
\nn
\nn
C^{\alpha\alpha}_{\text{III}}&\equiv A(x,\bar x)(z_1-z_3)(z_2-z_4)\bigg(\oint_{z=0}{dz\over 2\pi}z^{-m}{1\over  (z-z_1)^{1\over2} (z-z_2)^{1\over2} (z-z_3)^{1\over2} (z-z_4)^{1\over2} }\bigg)\nn 
&~~~\times\bigg(\oint_{z'=0}{dz'\over 2\pi}z'^{-n}{1\over (z'-z_1)^{1\over2} (z'-z_2)^{1\over2} (z'-z_3)^{1\over2} (z'-z_4)^{1\over2}}\bigg)
\end{align}
We observe a simple relation between terms $\text I$ and $\text{II}$ through pairwise interchanges of twist locations
\begin{equation}\label{Caa interchange}
C^{\alpha\alpha}_{\text{II}} = C^{\alpha\alpha}_{\text{I}}(z_1\leftrightarrow z_2, z_3\leftrightarrow z_4)
\end{equation}

\paragraph{Term I and II}

Let's begin by computing $C^{\alpha\alpha}_{\text I}$. Inserting the expansions around $z=0$ (\ref{expansion 0}) into the expression for $C^{\alpha\alpha}_{\text I}$ in (\ref{CI CII CIII}), we obtain the following expression
\begin{align}
C^{\alpha\alpha}_{\text I} =\, & -{1\over2}\sum_{k_i,k'_i\geq0}{}^{{1\over2}}C_{k_1}{}^{-{1\over2}}C_{k_2}{}^{{1\over2}}C_{k_3}{}^{-{1\over2}}C_{k_4}{}^{-{1\over2}}C_{k'_1}{}^{{1\over2}}C_{k'_2}{}^{-{1\over2}}C_{k'_3}{}^{{1\over2}}C_{k'_4}
\nn
&(-1)^{\sum k_i+\sum k'_i}z^{-(k_1+k'_1)}_1z^{-(k_2+k'_2)}_2z^{-(k_3+k'_3)}_3z^{-(k_4+k'_4)}_4
\nn 
&\oint_{z=0,\, |z|>|z'|}{dz\over 2\pi}z^{-m+\sum k_i}\oint_{z'=0}{dz'\over 2\pi}z'^{-n+\sum k'_i}{1\over(z-z')^2}
\end{align}
Focusing on the last line, we first perform the contour integral over $z'$, and then over $z$
\begin{align}
&(n-k'_1-k'_2-k'_3-k'_4)i\oint_{z=0}{dz\over 2\pi}z^{-m-n+k_1+k_2+k_3+k_4+k'_1+k'_2+k'_3+k'_4-1}\bigg|_{n-k'_1-k'_2-k'_3-k'_4>0}\nn
=\, & - (n-k'_1-k'_2-k'_3-k'_4)\delta_{m+n-k_1-k_2-k_3-k_4-k'_1-k'_2-k'_3-k'_4,0}\bigg|_{n-k'_1-k'_2-k'_3-k'_4>0}
\end{align}
This gives
\begin{align}\label{CI}
C^{\alpha\alpha}_{\text I} =& {1\over2}\sum_{k_i,k'_i\in D}{}^{{1\over2}}C_{k_1}{}^{-{1\over2}}C_{k_2}{}^{{1\over2}}C_{k_3}{}^{-{1\over2}}C_{k_4}{}^{-{1\over2}}C_{k'_1}{}^{{1\over2}}C_{k'_2}{}^{-{1\over2}}C_{k'_3}{}^{{1\over2}}C_{k'_4}
\nn
&(-1)^{\sum k_i+\sum k'_i}z^{-(k_1+k'_1)}_1z^{-(k_2+k'_2)}_2z^{-(k_3+k'_3)}_3z^{-(k_4+k'_4)}_4
(n-k'_1-k'_2-k'_3-k'_4)
\end{align}
with the sum taken over the region
\be
D= \big\{\ k_i,k'_i \geq 0\ ,\quad  m+n-\sum k_i-\sum k'_i=0\ , \quad n-\sum k'_i>0 \ \big\}
\ee
Following a similar procedure as before, this region can be written as
\begin{align}
k_3&\leq m+n-(k'_1+k'_2+k'_3+k'_4+k_1+k_2) \nn 
k_2&\leq m+n-(k'_1+k'_2+k'_3+k'_4+k_1) \nn
k_1&\leq m+n-(k'_1+k'_2+k'_3+k'_4) \nn
k'_4&\leq n-(k'_1+k_2'+k'_3+1) \nn   
k'_3&\leq n-(k'_1+k'_2+1)\nn
k'_2&\leq n-(k'_1+1)\nn
k'_1&\leq n-1    
\end{align}
which gives
\begin{align}\label{CI prime}
C^{\alpha\alpha}_{\text I} =\, & {1\over2}(-1)^{m+n}\sum_{k'_1=0}^{n-1}~\sum_{k'_2=0}^{n-k'_1-1}~\sum_{k'_3=0}^{n-k'_1-k'_2-1}~\sum_{k'_4=0}^{n-k'_1-k'_2-k'_3-1}~\sum_{k_1=0}^{m+n-k'_1-k'_2-k'_3-k'_4}\nn
&\sum_{k_2=0}^{m+n-k'_1-k'_2-k'_3-k'_4-k_1}~\sum_{k_3=0}^{m+n-k'_1-k'_2-k'_3-k'_4-k_1-k_2}(n-k'_1-k'_2-k'_3-k'_4)\nn 
&{}^{{1\over2}}C_{k_1}{}^{-{1\over2}}C_{k_2}{}^{{1\over2}}C_{k_3}{}^{-{1\over2}}C_{k_4}{}^{-{1\over2}}C_{k'_1}{}^{{1\over2}}C_{k'_2}{}^{-{1\over2}}C_{k'_3}{}^{{1\over2}}C_{k'_4}
\nn
&z^{-k_1-k'_1}_1z^{-k_2-k'_2}_2z^{-k_3-k'_3}_3z^{-m-n+k_1+k_2+k_3+k'_1+k'_2+k'_3}_4
\end{align}

For the second term in the contraction, $C^{\alpha\alpha}_{\text{II}}$, we can simply make the switches 
$z_1\leftrightarrow z_2$ and $z_3\leftrightarrow z_4$ in the above expression, as indicated in (\ref{Caa interchange}).

\paragraph{Term III}

Now, let's compute the third term in the contraction, $C^{\alpha\alpha}_{\text{III}}$. The derivation is similar to that of $f^{\alpha\alpha}_{\text{III}}$ and $f^{\alpha\bar\alpha}_{n,p}$, but with a different pole structure. Inserting (\ref{expansion 0}) into the expression for $C^{\alpha\alpha}_{\text{III}}$ (\ref{CI CII CIII}) yields
\begin{align}
C^{\alpha\alpha}_{\text{III}}=\ &{A(x,\bar x)(z_1-z_3)(z_2-z_4)\over z_1z_2z_3z_4}\nn 
&\times\bigg(\sum_{k_i\geq0}{}^{-{1\over2}}C_{k_1}{}^{-{1\over2}}C_{k_2}{}^{-{1\over2}}C_{k_3}{}^{-{1\over2}}C_{k_4}(-1)^{\sum k_i}z_1^{-k_1}z_2^{-k_2}z_3^{-k_3}z_4^{-k_4}\oint_{z=0}{dz\over 2\pi}z^{-m+\sum k_i}\bigg)
\nn 
&\times\bigg(\sum_{k'_i\geq0}{}^{-{1\over2}}C_{k'_1}{}^{-{1\over2}}C_{k'_2}{}^{-{1\over2}}C_{k'_3}{}^{-{1\over2}}C_{k'_4}(-1)^{\sum k'_i}z_1^{-k'_1}z_2^{-k'_2}z_3^{-k'_3}z_4^{-k'_4}\oint_{z'=0}{dz'\over 2\pi}z'^{-n+\sum k'_i}\bigg)
\end{align}
Performing the contour integrals of $z$ and $z'$ yields 
\begin{align}\label{CIII prime}
C^{\alpha\alpha}_{\text{III}}=\, &-{A(x,\bar x)(z_1-z_3)(z_2-z_4)\over z_1z_2z_3z_4}\nn 
&\times\bigg(\sum_{\substack{k_i\geq0\\m-\sum k_i-1=0}}{}^{-{1\over2}}C_{k_1}{}^{-{1\over2}}C_{k_2}{}^{-{1\over2}}C_{k_3}{}^{-{1\over2}}C_{k_4}(-1)^{\sum k_i}z_1^{-k_1}z_2^{-k_2}z_3^{-k_3}z_4^{-k_4}\bigg)
\nn 
&\times\bigg(\sum_{\substack{k'_i\geq 0\\n-\sum k'_i-1=0}}{}^{-{1\over2}}C_{k'_1}{}^{-{1\over2}}C_{k'_2}{}^{-{1\over2}}C_{k'_3}{}^{-{1\over2}}C_{k'_4}(-1)^{\sum k'_i}z_1^{-k'_1}z_2^{-k'_2}z_3^{-k'_3}z_4^{-k'_4}\bigg)
\end{align}
The region of the summation can be rewritten as follows
\begin{align} 
m-k_1-k_2-k_3-k_4-1=0&\implies k_4=m-k_1-k_2-k_3-1\nn
k_4\geq0&\implies m-k_1-k_2-1\geq k_3\nn
k_3\geq0&\implies m-k_1-1\geq k_2\nn
k_2\geq0&\implies m-1\geq k_1\nn
\nn
n-k'_1-k'_2-k'_3-k'_4-1=0&\implies k'_4=n-k'_1-k'_2-k'_3-1\nn
k'_4\geq0&\implies n-k'_1-k'_2-1\geq k'_3\nn
k'_3\geq0&\implies n-k'_1-1\geq k'_2\nn
k'_2\geq0&\implies n-1\geq k'_1
\end{align}
which gives
\begin{align}\label{CIII prime}
C^{\alpha\alpha}_{\text{III}}=\, &-{A(x,\bar x)(z_1-z_3)(z_2-z_4)\over z_1z_2z_3z_4}\nn 
&\times\bigg((-1)^m\sum_{k_1=0}^{m-1}\sum_{k_2=0}^{m-k_1-1}\sum_{k_3=0}^{m-k_1-k_2-1}{}^{-{1\over2}}C_{k_1}{}^{-{1\over2}}C_{k_2}{}^{-{1\over2}}C_{k_3}{}^{-{1\over2}}C_{m-k_1-k_2-k_3-1}\nn 
&\qquad z_1^{-k_1}z_2^{-k_2}z_3^{-k_3}z_4^{-m+k_1+k_2+k_3+1}\bigg)
\nn 
&\times\bigg((-1)^n\sum_{k'_1=0}^{n-1}\sum_{k'_2=0}^{n-k'_1-1}\sum_{k'_3=0}^{n-k'_1-k'_2-1}{}^{-{1\over2}}C_{k'_1}{}^{-{1\over2}}C_{k'_2}{}^{-{1\over2}}C_{k'_3}{}^{-{1\over2}}C_{n-k'_1-k'_2-k'_3-1}\nn
&\qquad z_1^{-k'_1}z_2^{-k'_2}z_3^{-k'_3}z_4^{-n+k'_1+k'_2+k'_3+1}\bigg)
\end{align}

\paragraph{Final expression for the left moving contraction}
By inserting the expressions (\ref{CI prime}), (\ref{Caa interchange}) and (\ref{CIII prime}) into (\ref{Caa}), we obtain
\begin{align}\label{Caa prime}
C^{\alpha\alpha}[m,n]=\, &{1\over2}(-1)^{m+n}\sum_{k'_1=0}^{n-1}~\sum_{k'_2=0}^{n-k'_1-1}~\sum_{k'_3=0}^{n-k'_1-k'_2-1}~\sum_{k'_4=0}^{n-k'_1-k'_2-k'_3-1}~\sum_{k_1=0}^{m+n-k'_1-k'_2-k'_3-k'_4}\nn
&\sum_{k_2=0}^{m+n-k'_1-k'_2-k'_3-k'_4-k_1}~\sum_{k_3=0}^{m+n-k'_1-k'_2-k'_3-k'_4-k_1-k_2}(n-k'_1-k'_2-k'_3-k'_4)\nn 
&{}^{-{1\over2}}C_{k'_1}{}^{{1\over2}}C_{k'_2}{}^{-{1\over2}}C_{k'_3}{}^{{1\over2}}C_{k'_4}{}^{{1\over2}}C_{k_1}{}^{-{1\over2}}C_{k_2}{}^{{1\over2}}C_{k_3}{}^{-{1\over2}}C_{m+n-k_1-k_2-k_3-k'_1-k'_2-k'_3-k'_4}
\nn
&~\times\Big(z^{-k_1-k'_1}_1z^{-k_2-k'_2}_2z^{-k_3-k'_3}_3z^{-m-n+k_1+k_2+k_3+k'_1+k'_2+k'_3}_4\nn 
&\qquad + z^{-k_1-k'_1}_2z^{-k_2-k'_2}_1z^{-k_3-k'_3}_4z^{-m-n+k_1+k_2+k_3+k'_1+k'_2+k'_3}_3\Big)
\nn 
&-{A(x,\bar x)(z_1-z_3)(z_2-z_4)\over z_1z_2z_3z_4}\nn 
&~~\times\bigg((-1)^m\sum_{k_1=0}^{m-1}\sum_{k_2=0}^{m-k_1-1}\sum_{k_3=0}^{m-k_1-k_2-1}{}^{-{1\over2}}C_{k_1}{}^{-{1\over2}}C_{k_2}{}^{-{1\over2}}C_{k_3}{}^{-{1\over2}}C_{m-k_1-k_2-k_3-1}\nn 
&~\qquad z_1^{-k_1}z_2^{-k_2}z_3^{-k_3}z_4^{-m+k_1+k_2+k_3+1}\bigg)
\nn 
&~~\times\bigg((-1)^n\sum_{k'_1=0}^{n-1}\sum_{k'_2=0}^{n-k'_1-1}\sum_{k'_3=0}^{n-k'_1-k'_2-1}{}^{-{1\over2}}C_{k'_1}{}^{-{1\over2}}C_{k'_2}{}^{-{1\over2}}C_{k'_3}{}^{-{1\over2}}C_{n-k'_1-k'_2-k'_3-1}\nn
&~\qquad z_1^{-k'_1}z_2^{-k'_2}z_3^{-k'_3}z_4^{-n+k'_1+k'_2+k'_3+1}\bigg)
\end{align}

The right moving contraction follows a similar derivation and can be obtained by taking the complex conjugate of the above expression
\be
C^{\bar\alpha\bar\alpha}[m,n] = (C^{\alpha\alpha}[m,n])^*
\ee
This means that we replace barred quantities with unbarred ones and vice versa, $z_i\leftrightarrow \bar z_i$. The final result is summarized in section \ref{results}.

\subsection{Left-right moving contraction}

For $C^{\alpha\bar\alpha}[m,n]$, the computation is very similar to that of $C^{\alpha\alpha}_{\text{III}}$ which is the third term of the left moving contraction. First recalling the expression for $C^{\alpha\bar\alpha}[m,n]$ recorded in (\ref{Caa Caabar}), we have
\begin{align}
C^{\alpha\bar\alpha}[m,n]&=-\oint_{z'=0}{dz'\over 2\pi}z'^{-m}\oint_{\bar z=0}{d\bar z\over 2\pi}\bar z^{-n}b(\bar z , z'; z_i, \bar z_i)
\end{align}
Inserting the correlation function for $b$ recorded in (\ref{g and b z plane}) into the above expression, we obtain
\begin{align}
&C^{\alpha\bar\alpha}[m,n]\nn
=\, &-B(x,\bar x)|z_1-z_3||z_2-z_4|\bigg(\oint_{z'=0}{dz'\over 2\pi}z'^{-m}{1\over (z'-z_1)^{1\over2} (z'-z_2)^{1\over2} (z'-z_3)^{1\over2} (z'-z_4)^{1\over2}}\bigg)
\nn
&\quad \times\bigg(\oint_{\bar z=0}{d\bar z\over 2\pi}\bar z^{-n}{1\over  (\bar z-\bar z_1)^{1\over2} (\bar z-\bar z_2)^{1\over2} (\bar z-\bar z_3)^{1\over2} (\bar z-\bar z_4)^{1\over2} }\bigg)
\end{align}
Notice that the above contour integral is similar to the integral in $C^{\alpha\alpha}_{\text{III}}$, as given in (\ref{CI CII CIII}).
We therefore obtain a final expression similar to (\ref{CIII prime})
\begin{align}
C^{\alpha\bar\alpha}[m,n]\label{Caabar final}
=\, &{B(x,\bar x)|z_1-z_3||z_2-z_4|\over|z_1z_2z_3z_4|}
\nn
&\times\bigg((-1)^m\sum_{k_1=0}^{m-1}\sum_{k_2=0}^{m-k_1-1}\sum_{k_3=0}^{m-k_1-k_2-1}{}^{-{1\over2}}C_{k_1}{}^{-{1\over2}}C_{k_2}{}^{-{1\over2}}C_{k_3}{}^{-{1\over2}}C_{m-k_1-k_2-k_3-1}\nn 
&\qquad z_1^{-k_1}z_2^{-k_2}z_3^{-k_3}z_4^{-m+k_1+k_2+k_3+1}\bigg)
\nn 
&\times\bigg((-1)^n\sum_{k'_1=0}^{n-1}\sum_{k'_2=0}^{n-k'_1-1}\sum_{k'_3=0}^{n-k'_1-k'_2-1}{}^{-{1\over2}}C_{k'_1}{}^{-{1\over2}}C_{k'_2}{}^{-{1\over2}}C_{k'_3}{}^{-{1\over2}}C_{n-k'_1-k'_2-k'_3-1}\nn
&\qquad \bar z_1^{-k'_1}\bar z_2^{-k'_2}\bar z_3^{-k'_3}\bar z_4^{-n+k'_1+k'_2+k'_3+1}\bigg)
\end{align}

\subsection{Relating contraction and pair creation}

Since both the initial and final states belong to the untwisted sector, a time-reversal symmetry exists that relates contraction and pair creation. In this section, we derive these relations and establish a similar relation for the propagation coefficients. These relations can serve as a cross-check for our results collected in section \ref{results}.

The contraction coefficients are given by (\ref{Caa Caabar z})
\begin{align}\label{Caa Caabar w II}
C^{\alpha\alpha}_{m,n}(z_i,\bar z_i)&= {\langle 0|  \prod_{i}\sigma(z_i,\bar z_i)\, \a_{-m}\a_{-n} |0\rangle\over \langle0| \prod_{i}\sigma(z_i,\bar z_i)| 0\rangle}
\nn
C^{\alpha\bar\alpha}_{m,n}(z_i,\bar z_i)&= {\langle 0|  \prod_{i}\sigma(z_i,\bar z_i)\, \a_{-m}\bar \a_{-n} |0\rangle\over \langle0| \prod_{i}\sigma(z_i,\bar z_i)| 0\rangle}
\end{align}
Here, we keep the discussion general, allowing the number and order of twist operators to be arbitrary, rather than restricting to the specific case of four twist-2 operators considered in this paper. 
To relate these coefficients to pair creation, we use the Hermitian conjugate of the twist operator
\be\label{conj s}
(\sigma(z,\bar z))^\dagger = \bar z^{-2h}z^{-2\bar h}\sigma(1/{\bar z},1/{ z})
\ee
where $h$ and $\bar h$ are the left and right moving dimensions of $\sigma$, respectively.
By taking the Hermitian conjugate of the contraction coefficients in (\ref{Caa Caabar w II}), we obtain
\begin{align}\label{C gamma}
(C^{\alpha\alpha}_{m,n}(z_i,\bar z_i))^*&= {\langle 0| \a_{n}\a_{m} \prod_{i}\sigma({1}/{\bar z_i},{1}/{ z_i})  |0\rangle\over \langle0| \prod_{i}\sigma({1}/{\bar z_i},{1}/{ z_i})| 0\rangle}
= 2mn\,\gamma_{m,n}({1}/{\bar z_i},{1}/{ z_i}) 
\nn
(C^{\alpha\bar\alpha}_{m,n}(z_i,\bar z_i))^*&= {\langle 0| \bar \a_{n}\a_{m} \prod_{i}\sigma({1}/{\bar z_i},{1}/{ z_i})  |0\rangle\over \langle0| \prod_{i}\sigma({1}/{\bar z_i},{1}/{ z_i})| 0\rangle}
=mn\,\beta_{m,n}({1}/{\bar z_i},{1}/{ z_i})
\end{align}
Notice that the prefactor in (\ref{conj s}) cancels between the numerator and denominator. Taking the complex conjugate of these expressions, we find the relation 
\begin{equation}\label{C gamma}
    C^{\alpha\alpha}_{m,n}(z_i,\bar z_i)  = 2mn\,\gamma_{m,n}(1/z_i,1/\bar z_i) \ ,\quad C^{\alpha\bar\alpha}_{m,n}(z_i,\bar z_i)  = mn\,\beta_{m,n}(1/z_i,1/\bar z_i)
\end{equation}
A similar relation can be derived for the propagation coefficients. Recall that the propagation coefficients are given by 
\begin{align}\label{faa relation}
f^{\alpha\alpha}_{n,p}(z_i,\bar z_i)&
= {1\over p}{\langle 0| \a_{p} \prod_{i}\sigma(z_i,\bar z_i)\, \a_{-n} |0\rangle\over \langle0| \prod_{i}\sigma(z_i,\bar z_i)| 0\rangle}\nn
f^{\alpha\bar \alpha}_{n,p}(z_i,\bar z_i)&
= {1\over p}{\langle 0| \bar \a_{p} \prod_{i}\sigma(z_i,\bar z_i)\, \a_{-n} |0\rangle\over \langle0| \prod_{i}\sigma(z_i,\bar z_i)| 0\rangle}
\end{align} 
Applying Hermitian conjugation to these expressions gives
\begin{align}
(f^{\alpha\alpha}_{n,p}(z_i,\bar z_i))^*&
= {1\over p}{\langle 0| \a_{n} \prod_{i}\sigma(1/\bar z_i,1/ z_i)\, \a_{-p} |0\rangle\over \langle0| \prod_{i}\sigma(1/\bar z_i,1/ z_i)| 0\rangle}=\frac{n}{p}f^{\alpha\alpha}_{p,n}(1/\bar z_i,1/ z_i)
\nn
(f^{\alpha\bar \alpha}_{n,p}(z_i,\bar z_i))^*&
= {1\over p}{\langle 0| \bar \a_{n} \prod_{i}\sigma(1/\bar z_i,1/ z_i)\, \a_{-p} |0\rangle\over \langle0| \prod_{i}\sigma(1/\bar z_i,1/ z_i)| 0\rangle}=\frac{n}{p}f^{\bar\alpha\alpha}_{p,n}(1/\bar z_i,1/ z_i)
\end{align}
Taking the complex conjugate of these expressions yields
\begin{align}\label{f relation}
f^{\alpha\alpha}_{n,p}(z_i,\bar z_i)
=\frac{n}{p}f^{\alpha\alpha}_{p,n}(1/ z_i,1/ \bar z_i)\ ,
\qquad
f^{\alpha\bar \alpha}_{n,p}(z_i,\bar z_i)
=\frac{n}{p}f^{\bar\alpha\alpha}_{p,n}(1/ z_i,1/ \bar z_i)
\end{align}

\section{Numerical analysis}\label{analysis}

In this section, we analyze the propagation and contraction coefficients numerically. 
We will work in the Lorentzian signature on the cylinder coordinate $(t=-i\tau,\sigma)$, as defined in (\ref{cylinder coord}). Notice that $\sigma$ here is the circular coordinate of the base space and not the twist operator.

First, we note that the propagation and contraction coefficients are invariant under any permutation of the twist locations $z_i\leftrightarrow z_j$ ($w_i \leftrightarrow w_j$) for $i,j = 1,2,3,4$. This symmetry arises because the correlation functions from which they are derived, $g$ and $b$, are invariant under such twist interchanges, as reviewed in subsection (\ref{sec pc}). We have also numerically verified this symmetry.

By direct inspection of the expressions collected in section \ref{results}, as well as through numerical testing, we find that the propagation coefficients exhibit the following anti-periodicity under shifts by $2\pi$ in any one of the twist locations
\begin{equation}\label{antiperiodicity}
    f^{\alpha\alpha}_{n,p}(t_i + 2\pi)=-f^{\alpha\alpha}_{n,p}(t_i), \quad f^{\alpha\bar\alpha}_{n,p}(t_i + 2\pi)=-f^{\alpha\bar\alpha}_{n,p}(t_i)
\end{equation}
In contrast, the contraction coefficients are periodic in $2\pi$, similar to the pair creation coefficients studied in \cite{Guo:2024edj}
\begin{equation}
    C^{\alpha\alpha}[n,p](t_i + 2\pi)=C^{\alpha\alpha}[n,p](t_i),\quad C^{\alpha\bar\alpha}[n,p](t_i + 2\pi)=C^{\alpha\bar\alpha}[n,p](t_i)
\end{equation}

We note that in our analysis we fix the twist operators to be located at the same point along the circle which we fix to be $\sigma_i=0$, and we study separations in the time direction. The reason that there is a shift by a minus sign for propagation is because the modes utilized in our analysis, we recall, are written as an antisymmetric combination of copy 1 and copy 2 modes since we've taken the number of copies of the CFT to be $N=2$. Therefore a $2\pi$ shift in $t_i$ for any one twist operator maps copy 1 into copy 2 and copy 2 into copy 1. Since propagation, $f^{\alpha\alpha},f^{\alpha\bar\alpha}$, multiplies a linear combination of modes in the copy 1 - copy 2 basis this gives an overall minus sign unlike the case for pair creation, $\gamma,\beta$, which multiplies a bilinear combination of modes in the copy 1 - copy 2 basis and contraction, $C^{\alpha\alpha},C^{\alpha\bar\alpha}$ which doesn't multiply any modes at all.

\begin{figure}
\centering
        \includegraphics[width=8cm]{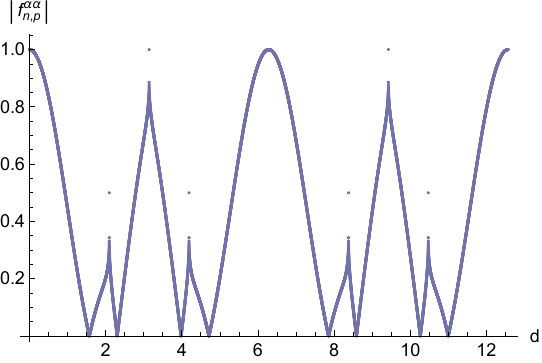}
        \includegraphics[width=8cm]{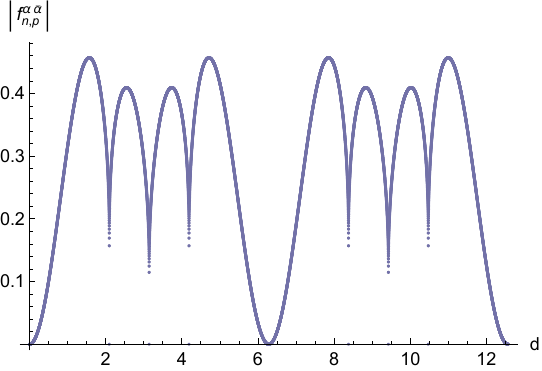}
\caption{$|f^{\alpha\alpha}_{n,p}|$ (left) and $|f^{\alpha\bar\alpha}_{n,p}|$ (right) vs. distance $d$ between equally spaced twist operators in time, with $\sigma_1=\sigma_2=\sigma_3=\sigma_4=0$, and $t_1=0$, $t_2=d$, $t_3=2d$, $t_4=3d$ with a range of $d$ from $0$ to $4\pi$ for $n=p=1$.}
\label{fign1p10to4pi}
\end{figure}

In the following, we focus on the propagation effects, as the contraction effects are very similar to pair creation, a relationship we proved explicitly in (\ref{C gamma}). 
Let us consider a configuration in which all twist operators are equally spaced in time, with the separation denoted by $d$, and positioned at the same spatial location, $\sigma_i = 0$. Due to the twist interchange symmetry and the periodicity, the plot exhibits reflective symmetry about $d = \pi$ and periodicity of $2\pi$, as shown in fig.\ref{fign1p10to4pi}.

\begin{figure}
\centering
        \includegraphics[width=8cm]{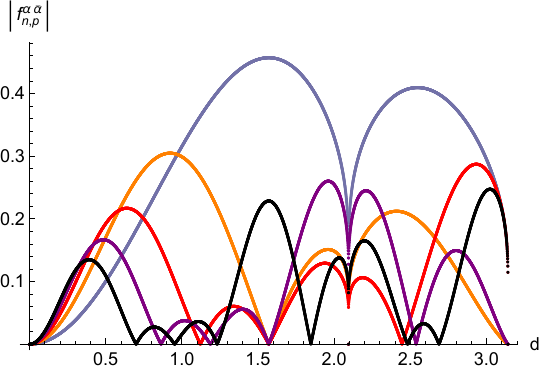}
\caption{$|f^{\alpha\bar\alpha}_{n,p}|$  vs. distance $d$ between equally spaced twist operators in time for initial mode $n=1$ and final modes $p=1$ (blue), $2$ (orange), $3$ (red), $4$ (purple), $5$ (black) for $\sigma_1=\sigma_2=\sigma_3=\sigma_4=0,$ and $t_1=0$, $t_2=d$, $t_3=2d$, $t_4=3d$ for $0\leq d\leq\pi$.}\label{fign1p12345}
\end{figure}

\begin{figure}
\centering
        \includegraphics[width=8.003cm]{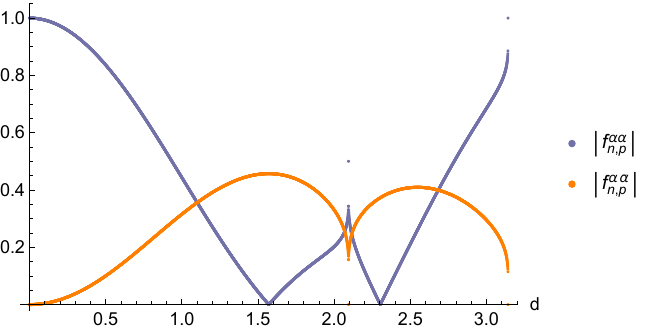}
        \includegraphics[width=8.003cm]{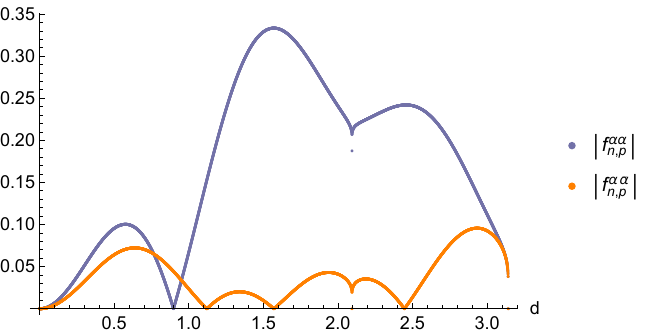}
\caption{$|f^{\alpha\alpha}_{n,p}|$ (blue) and $|f^{\alpha\bar\alpha}_{n,p}|$ (orange) vs. distance $d$ between equally spaced twist operators in time for initial mode $n=1$ and final mode $p=1$ (left) and $p=3$ (right), for $\sigma_1=\sigma_2=\sigma_3=\sigma_4=0,$ and $t_1=0$, $t_2=d$, $t_3=2d$, $t_4=3d$ for $0\leq d\leq\pi$.}\label{fign1p1p3}
\end{figure}

Let us note several other features of propagation from fig.\ref{fign1p10to4pi}. When all four twist operators coincide, they can be replaced by the identity operator since there is no effect at all. In this case the initial mode simply propagates unaltered on the base space as time evolves. Therefore, propagation, characterizing the mode transition $\alpha\to\alpha$, is given by $|f^{\alpha\alpha}_{n,n}|=1$ and $|f^{\alpha\alpha}_{n,p}|=0$ for $p\neq n$ with the transition $\alpha\to\bar\alpha$ yielding $|f^{\alpha\bar\alpha}_{n,p}|=0$, as expected, a feature which is also explicitly highlighted in fig.\ref{fign1p1p3} where the left plot is for equal energy between initial and final modes, $n=p=1$, and the right plot is for differing energies between the initial and final modes, $n=1,p=3$. As the twist operators are separated in time, the probability to transition into modes with different energies and modes in different sectors becomes nonzero. At $d={2\pi\over3}$ the first and fourth twist operators are separated by $2\pi$ which, because of periodicity conditions (\ref{antiperiodicity}), is equivalent to the first and fourth twist operators coinciding, which yields the two twist configuration. At this separation mixed propagation vanishes and standard propagation reduces to the two twist result. Increasing the separation to $d=\pi$ corresponds to the first and third twist operators being separated by 
$2\pi$ and the second and fourth twist operators being separated by $2\pi$. Which, again through (\ref{antiperiodicity}), is equivalent to the first and third twist operators coinciding and the second and fourth twist operators coinciding with the two pairs being separated by a distance of $\pi$.  
For this configuration propagation for $\alpha\to\alpha$ again returns to a value of $|f^{\alpha\alpha}_{n,n}|=1$, $|f^{\alpha\alpha}_{n,p}|=0$ for $p\neq n$ and for $\alpha\to\bar\alpha$ to a value of $|f^{\alpha\bar\alpha}_{n,p}|=0$. These features are demonstrated numerically in fig.\ref{fign1p10to4pi}. 
Fig.\ref{fign1p12345} shows the behavior of the left-right mixed propagation $|f^{\alpha\bar\alpha}_{n,p}|$ as a function of the equal time separation, $d$, between consecutive twist operators. It takes the largest value at $n=p=1$ and $d={\pi\over2}$, which is the widest separation among the four twist operators within a periodic $2\pi$ interval.
This aligns with the role of twist operators, which join and split copies of the CFTs. In time order, the first and third twist operators join two untwisted CFT copies into a twisted copy, while the second and fourth split it back into two untwisted copies. As a result, consecutive twist operators tend to cancel each other's effects when they are close. 
Therefore, the maximum effect arises at $d=\frac{\pi}{2}$, where the twist operators are most widely separated.
For higher energy modes (shorter wavelengths), smaller separations dominate, resulting in stronger cancellations and smaller propagation coefficients. These behaviors are similar to the left-right mixed pair creation studied in \cite{Guo:2024edj}.

\begin{figure}
\centering
        \includegraphics[width=8cm]{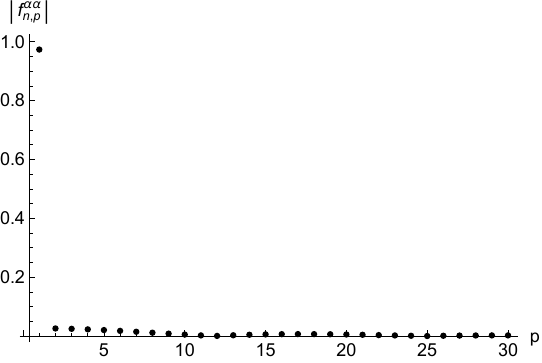}
        \includegraphics[width=8cm]{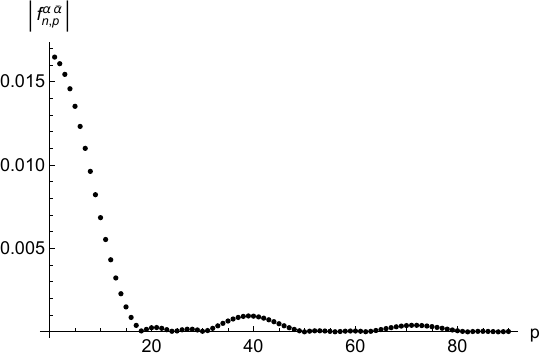}
\caption{$|f^{\alpha\alpha}_{n,p}|$ (left) and $|f^{\alpha\bar\alpha}_{n,p}|$ (right) vs. $p$, the final mode number, with fixed initial mode number $n=1$ for equally spaced twist operators for $\sigma_1=\sigma_2=\sigma_3=\sigma_4=0,$ and $t_1=0$, $t_2={\pi\over16}$, $t_3={\pi\over8}$, $t_4={3\pi\over16}$.}\label{fign1p}
\end{figure}

As can be seen in fig.\ref{fign1p}, for small time separations between twist operators we see that propagation $|f^{\alpha\bar\alpha}_{n,p}|$ is highly peaked when the final energy is equal to the initial energy which in this case is, $n=p=1$, and sharply decays for $p\neq n$. For the same separation the value of mixed propagation, $|f^{\alpha\bar\alpha}_{n,p}|$, is small for any value of $p$ with a decaying envelope starting from $p=1$. This behavior is expected, again demonstrating the cancellation effects by twist operators which act close together. 

\begin{figure}
\centering
        \includegraphics[width=8cm]{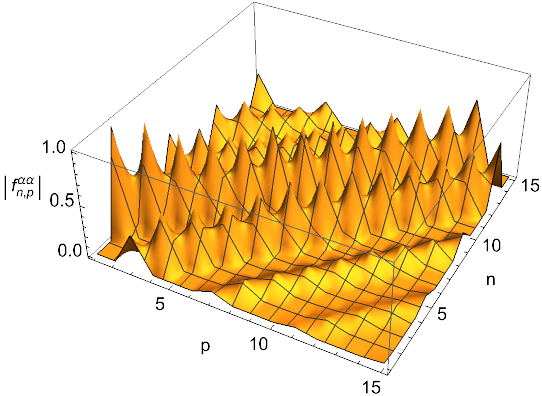}
        \includegraphics[width=8cm]{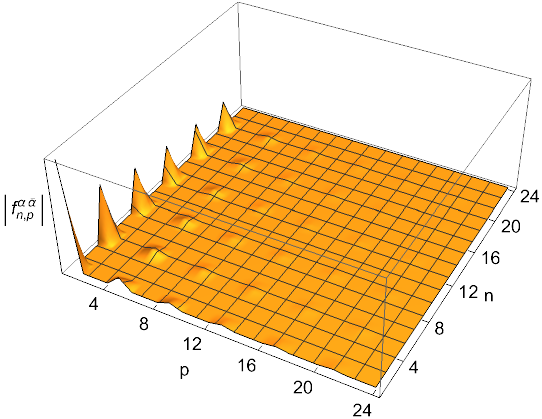}
\caption{$|f^{\alpha\alpha}_{n,p}|$ (left), $|f^{\alpha\bar{\alpha}}_{n,p}|$ (right)  vs. $n,p$  for  $\sigma_1=\sigma_2=\sigma_3=\sigma_4=0,$ and $t_1=0$, $t_2={\pi\over2}$, $t_3=\pi$, $t_4={3\pi\over2}$. Although $n,p$ are integers, smooth interpolation is used for better visualization.}\label{fig3D}
\end{figure}

Fig.\ref{fig3D} shows values of $|f^{\alpha\bar\alpha}_{n,p}|$ and $|f^{\alpha\bar\alpha}_{n,p}|$ vs. $n,p$ for maximum equal separation of consecutive twist operators in time, $d={\pi\over2}$. We see that $|f^{\alpha\alpha}_{n,p}|$ strongly oscillates for various values of $n$ and $p$, giving zero of $n=p$ which is consistent with expected behavior. For mixed propagation $|f^{\alpha\bar\alpha}_{n,p}|$ we see that it peaks at $n=p=1$ and yields decaying local maxima for larger values of $n$ with $p=1$. While the lowest mode for both $n,p$ yields the highest transition probability, for higher energy initial left moving modes, $n>1$, for which $|f^{\alpha\bar\alpha}_{n,p}|$ is nonzero, for each fixed $n$, the largest transition is to a right moving mode with the lowest energy, $p=1$. This nicely captures the behavior of nontrivial monodromy as a low energy effect.

\newpage

\section{Some limits}\label{limits}

The following question is crucial for understanding the effect of multiple twist operators: Can we reproduce the effect of multiple twist operators by summing over intermediate states and using the effects of one twist or fewer twists? If so, it would simplify the computation of multiple twist effects. In this section, we will investigate this question by analyzing the effect of four-twist operators in two different limits.

The first is the continuum limit, where the energy corresponding to mode numbers becomes very large. In \cite{Carson:2014xwa}, it is assumed that in this limit, the effects of multiple twists can be reproduced by combining the effects of single twists. We will find that our results strongly support this assumption.
The second limit is the large separation limit, where the distance between twists becomes very large while the mode numbers remain relatively low. In this case, we find that nontrivial monodromy effects still persist, even though the twist operators are well-separated. 

\subsection{Continuum Limit}

In \cite{Carson:2014ena,Carson:2015ohj,Carson:2016cjj}, the large-energy (or continuum) limit was taken to obtain approximate expressions for one- and two-twist effects, including pair creation and propagation. This involved considering mode numbers which were $\gg1$. For the single-twist case, the continuum limit is more straightforward to apply, as the expressions can be obtained in closed analytic form. For the propagation of a single twist, we have
\be\label{one conti}
f^{B(1)}_{np} \approx 
\begin{cases}
\frac{i}{(n-p)\pi}\sqrt{\frac{n}{p}}\, e^{w_0(p-n)}\qquad& n\neq p\\
\quad \frac{1}{2}& n = p
\end{cases}
\ee
where $n,p\gg 1$. In this process, two untwisted copies join into a twist-2 copy. Here $(1)$ labels an initial copy, and $B$ denotes the bosonic field. For more details, see \cite{Carson:2014yxa,Carson:2014xwa}. The parameter $w_0$ is the location of the twist operator on the cylinder. For pair creation, the continuum limit is given by
\be 
\gamma^B_{mn}\approx{1\over2\pi}{1\over\sqrt{mn}}{1\over m+n}e^{w_0(m+n)}
\ee

For the two-twist case, however, the situation is a bit more complicated. The exact expressions are not simple analytic functions as in the single twist case but are instead given by finite sums involving binomial coefficients and powers of twist locations. 
Furthermore, since the presence of two or more twist operators introduces twist separations, the expressions for the corresponding twist effects have an oscillatory behavior governed by this separation. The separation between the two twists in Lorentzian signature is denoted by $\Delta w=i(\Delta t+ \Delta \sigma)$. The continuum limit for propagation in the two-twist case is given by
\begin{equation}\label{2twistf}
 f^{B,(1)(1')}_{np}\approx\begin{cases}
   {1\over(n-p)\pi}\sqrt{n\over p}\sin\big((n-p){\Delta w\over2\pi i}\big)\text{sgn}\big({\Delta w\over 2\pi i}-1\big),\qquad&n\neq p\\
   \big|{\Delta w\over2\pi i}-1\big|,\qquad&n=p
    \end{cases}
\end{equation}
where $(1)$ denotes initial copy 1 and $(1')$ denotes final copy 1. For more details, see \cite{Carson:2016cjj}.
We observe that for $n\neq p$, the amplitude ${1\over(n-p)\pi}\sqrt{n\over p}$ of the oscillatory function matches that of the single-twist case in (\ref{one conti}).

For pair creation, the oscillatory function has not yet been determined explicitly but its amplitude has and matches that of the single-twist case \cite{Carson:2015ohj}
\begin{equation}
\gamma^{B,(1)(1)}_{mn} \sim {1\over\sqrt{mn}(m+n)}
\end{equation}
where both modes correspond to copy-1 excitations in the final state. 
Since the amplitude of the oscillatory function for both pair creation and propagation remains the same in the one- and two-twist cases, it has been conjectured that for an arbitrary number of twist operators they should similarly display this same amplitude \cite{Carson:2016cjj}. The oscillating factor however is expected to become increasingly more complicated as the number of twist operators increases.

Turning to the four-twist results computed here and in \cite{Guo:2024edj}, we note that the pair creation and propagation expressions in (\ref{gamma final collect}) and (\ref{f aa collect}) share similarities with the two-twist case, involving finite sums but with an additional nontrivial monodromy term. The oscillatory behavior of the four-twist expressions is more complicated due to the increased number of possible combinations of twist separations. To simplify, we consider a configuration where all four twist operators are at the same spatial location, $\sigma_1=\sigma_2=\sigma_3=\sigma_4=0$, and are equally spaced in time at $t_1=0$, $t_2=d$, $t_3= 2d$ and $t_4= 3d$. We will study how the oscillatory behavior depends on $d$.

\begin{figure}
\centering
        \includegraphics[width=12cm]{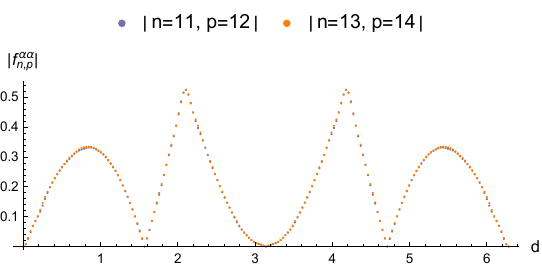}
\caption{The plot of $|f^{\alpha\alpha}_{n,p}|$ vs. $d$ for  $\sigma_1=\sigma_2=\sigma_3=\sigma_4=0$, and $t_1=0$, $t_2=d$, $t_3= 2d$, $t_4= 3d$, over the range $0\leq d\leq 2\pi$ for $n=11,p=12$ (blue) and $n=13,p=14$ (orange).}\label{fconpminusn1}
\end{figure}

\begin{figure}
\centering
        \includegraphics[width=12cm]{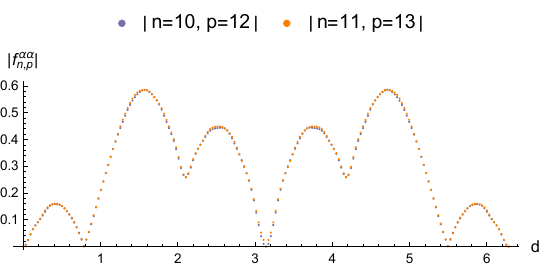}
\caption{The plot of $|f^{\alpha\alpha}_{n,p}|$ vs. $d$ for  $\sigma_1=\sigma_2=\sigma_3=\sigma_4=0$, and $t_1=0$, $t_2=d$, $t_3= 2d$, $t_4= 3d$, over the range $0\leq d\leq 2\pi$ for $n=10,p=12$ (blue) and $n=11,p=13$ (orange).}\label{fconpminusn2}
\end{figure}

For propagation, we plot the energy differences for $p-n=1$ in fig.\ref{fconpminusn1} and for $p-n=2$ in fig.\ref{fconpminusn2}. Unlike the one- and two-twist continuum limits, which exhibit an analytic form that clearly separates the amplitude from the oscillatory function as illustrated in (\ref{one conti}) and (\ref{2twistf}), the four-twist case does not display such a clean separation. Nevertheless, we observe that in the continuum limit ($n,p\gg 1$) with fixed $n-p$, the propagation approaches a fixed profile, as seen in fig.\ref{fconpminusn1} and fig.\ref{fconpminusn2}, where the two curves match precisely. This behavior is also seen in the one- and two-twist cases discussed earlier. Similarly, for pair creation in the continuum limit ($m,n\gg 1$) with fixed $m+n$, the profile approaches a fixed form, as shown in fig.\ref{gamcon}.
\begin{figure}
\centering
        \includegraphics[width=12cm]{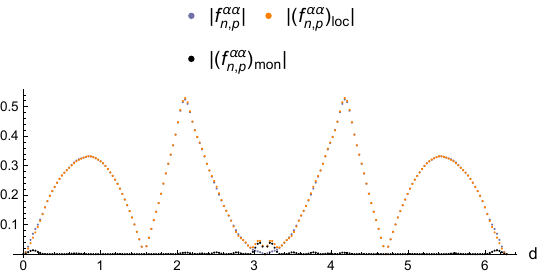}
\caption{The plot of $|f^{\alpha\alpha}_{n,p}|$ (blue), $|(f^{\alpha\alpha}_{n,p})_{\text{local}}|$ (orange), $|(f^{\alpha\alpha}_{n,p})_{\text{mono}}|$ (black) vs. $d$ for  $\sigma_1=\sigma_2=\sigma_3=\sigma_4=0$, and $t_1=0$, $t_2=d$, $t_3= 2d$, $t_4= 3d$, over the range $0\leq d\leq 2\pi$ for $n=11,p=12$.}\label{fcomp}
\end{figure}
\begin{figure}
\centering
        \includegraphics[width=12cm]{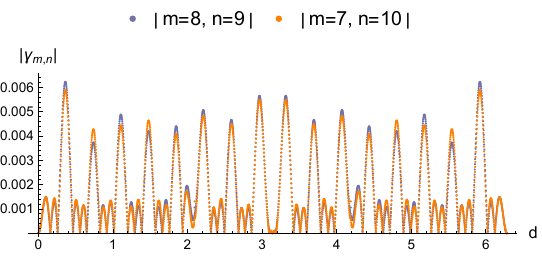}
\caption{The plot of $|\g_{m,n}|$ vs. $d$ for  $\sigma_1=\sigma_2=\sigma_3=\sigma_4=0$, and $t_1=0$, $t_2=d$, $t_3= 2d$, $t_4= 3d$, over the range $0\leq d\leq 2\pi$ for $m=8,n=9$ (blue) and $m=7,n=10$ (orange).}\label{gamcon}
\end{figure}  
\begin{figure}
\centering
        \includegraphics[width=12cm]{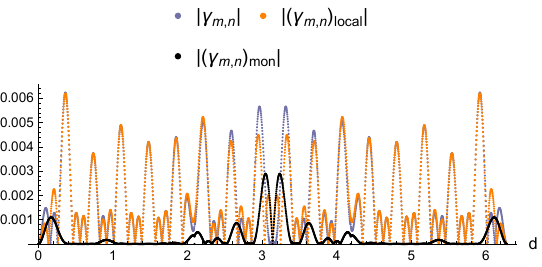}
\caption{The plot of $|\g_{m,n}|$ (blue), $|(\g_{m,n})_{\text{local}}|$ (blue) and $|(\g_{m,n})_{\text{mon}}|$ vs. $d$ for  $\sigma_1=\sigma_2=\sigma_3=\sigma_4=0$, and $t_1=0$, $t_2=d$, $t_3= 2d$, $t_4= 3d$, over the range $0\leq d\leq 2\pi$ for $m=8,n=9$.}\label{gamcomp}
\end{figure}

Returning to the question of whether the effect of four twist operators can be produced by combining the effects of two pairs of two twists. The four-twist effects include nontrivial monodromy terms, such as those with coefficients $A(x,\bar x)$ and $B(x,\bar x)$ in the expressions summarized in section \ref{results}. 
In the continuum limit, these nontrivial monodromy terms decay much more rapidly than the local terms. 
This rapid falloff at higher energies indicates that these terms are important only at low energies, as they arise from nontrivial cycles on the torus covering surface, with their size determined by the separation of the twist operators. We demonstrate this behavior by comparing the full result, the local term and the monodromy term both for pair creation, $\gamma_{m,n}$ for energies $m=8,n=9$, shown in fig.\ref{gamcomp}, and for propagation, $f^{\alpha\alpha}_{n,p}$ for energies $n=11,p=12$ shown in fig.\ref{fcomp}, where each effect is plotted as function of equal time twist separation $d$. We note that when the twists are sufficiently separated, the monodromy terms decay quickly. For separations in which the twist operators are close together, either directly or through (anti)periodic shifts of $2\pi$, the monodromy terms have nontrivial behavior and may have values comparable to those of the local terms. This is because short separations correspond to energies comparable to those in the high energy limit which, for these separations, is no longer reliable.   
The negligibility of the monodromy terms at higher energies strongly supports the assumption that the effect of multiple twist operators at high energy can be reproduced by combining the effects of fewer twist operators.

\subsection{Small and large separation limit}

We now consider the small and large separation limits of twist effects for four twist operators. To do this, we make the following transformations 
\begin{equation}\label{transf}
c_1\equiv z_1-z_2,\quad d\equiv z_2-z_3,  \quad c_2\equiv z_3-z_4,\quad s=\frac{1}{4}(z_1+z_2+z_3+z_4)
\end{equation}
We will focus on the case where $c_1$, $c_2$ and $s$ are held fixed, while considering the limits $d\to 0$ or $d\to \infty$.

\subsubsection*{\underline{Small separation limit $d=z_2-z_3\to0$}}

Let us consider the limit where $z_2\to z_3$ or $d\to0$ with $c_1,c_2$ and $s$ fixed. 
First we analyze the monodromy functions $A(z_i,\bar z_i)$ and $B(z_i,\bar z_i)$, from (\ref{collect AB})
\begin{align}\label{A app}
A(z_i,\bar z_i) =& -(z_1-z_3)(z_2-z_4)A(x,\bar x)
\nn
=&-(z_1-z_3)(z_2-z_4)\nn
&\times{(E(x) - (1-x)K(x))\bar K(1-\bar x)-( E(1-x) - xK(1-x) )\bar K(\bar x) \over2 (K(x)\bar K(1-\bar x)+  K(1-x)\bar K(\bar x))}
\end{align}
and
\begin{align}\label{B app}
B(z_i,\bar z_i) &= |z_1-z_3||z_2-z_4|B(x,\bar x)
\nn
&={|z_1-z_3||z_2-z_4|\pi\over 4(K(x)\bar K(1-\bar x)+  K(1-x)\bar K(\bar x))}
\end{align}
By inserting (\ref{transf}) into (\ref{A app}) and (\ref{B app}), and recalling the relation
\begin{equation}
    x ={(z_1-z_2)(z_3-z_4)\over(z_1-z_3)(z_2-z_4)}
\end{equation}
we can expand around $d\sim0$, keeping only the leading order terms. The resulting behavior as $d\to0$ is
\begin{align}\label{dto0}
 A(z_i,\bar z_i)&\approx  {c_1c_2\over2\log|d|}\to0\nn
B(z_i,\bar z_i)&\approx  -{|c_1c_2|\over2\log|d|}\to0
\end{align}
This is consistent with the expectation that the monodromy terms vanish as two twist operators are brought together. In this limit, the four twist configuration reduces to the two twist configuration, which does not contain nontrivial monodromy terms. For pair creation (\ref{gamma final collect}), (\ref{beta final}) and propagation (\ref{f aa collect}), (\ref{f aabar collect}) for the lowest mode numbers, we find the following limits as $d\to0$
\begin{align}
\gamma_{1,1}&\approx {1\over16}(c_1+c_2)^2+{1\over4}{c_1c_2\over \log|d|}\to {1\over16}(c_1+c_2)^2
={1\over16}(z_1-z_2+z_3-z_4)^2
\nn
\beta_{1,1}&\approx -{|c_1c_2|\over2\log|d|}\to0
\nn
f^{\alpha\alpha}_{1,1}&\approx{1\over2}\bigg(\sqrt{3c_1+c_2+4s\over -c_1-3c_2+4s}+\sqrt{-c_1-3c_2+4s\over3c_1+c_2+4s}\bigg)={1\over2}\bigg(\sqrt{z_1\over z_4}+\sqrt{z_4\over z_1}\bigg)
\nn
f^{\alpha\bar\alpha}_{1,1}&\approx  -{8|c_1c_2|\over\sqrt{(-c_1-3c_2+4s)(3c_1+c_2+4s)(-c_1+c_2+4s)^2}\log |d|}\to0
\end{align}
In the limit where two of the four twists come together, the left-moving effects remain, approaching the two-twist result \cite{Carson:2015ohj,Carson:2016cjj}. The left-right mixed effects vanish because there is no nontrivial monodromy in the two-twist case. These results have also been proven more generally in \cite{Guo:2024edj}.

\subsubsection*{\underline{Large separation limit $d=z_2-z_3\to\infty$}}

Let us now consider the limit $d=z_2-z_3\to\infty$ again with $c_1,c_2,s$ held fixed. 
Expanding (\ref{A app}) and (\ref{B app}), around $d\sim\infty$ and explicitly writing the leading order terms, we find the limits for the two monodromy functions to be
\begin{align}
    A(z_i,\bar z_i) &\approx {d^2\over4\log|d|}\to\infty,
\nn
B(z_i,\bar z_i)&\approx {|d|^2\over4\log|d|}\to\infty
\end{align}
Considering the lowest mode number contributions for pair creation (\ref{gamma final collect}), (\ref{beta final}) and propagation (\ref{f aa collect}), (\ref{f aabar collect}), in the limit that $d\to\infty$, we find that
\begin{align}
\gamma_{1,1}&\approx{d^2\over 8\log|d|}\to\infty
\nn
\beta_{1,1}&\approx{|d|^2\over4\log|d|} \to\infty
\nn
f^{\alpha\alpha}_{1,1}&\approx 1
\nn
 f^{\alpha\bar\alpha}_{1,1}&\approx{\bar d\over d\log\bar d} \to0
\end{align}
For $\gamma_{1,1}$, in the large $d$ limit, the dominant contribution is naively expected to come from the vacuum state in the intermediate states between the second and third twists, scaling proportionally to $d^2$. The additional factor of $\frac{1}{\log|d|}$ arises from the nontrivial monodromy effects, which cannot be simply reproduced by summing over the excited intermediate states, as they contribute only lower powers of $d$. In contrast, for $f^{\alpha\alpha}_{1,1}$, the dominant contribution comes from the first excited states $\alpha_{-1}$  between the second and third twists (since the vacuum contribution vanishes, as the initial state contains only one boson, which cannot turn into a vacuum by the first two twists). This reproduces the same scaling as in the above result. There is no nontrivial monodromy effect at leading order in the large $d$ limit. The discrepancy in the leading behavior of $\gamma$ and $f$ may arise because, in the case of pair creaton, the vacuum contributes, while in propagation, it does not. We conjecture that the following is generally true: for pair creation (and contraction), there is a nontrivial monodromy effect at leading order in the large $d$ limit, whereas there is none in propagation. It will be interesting to explore this further. 

\section{Collecting Results}\label{results}

Here we collect the expressions for all the effects of the twist operators: pair creation, as computed in \cite{Guo:2024edj}, along with propagation and contraction, which are computed in this paper.

In the following the absolute value is given by $|z|=\sqrt{z\bar z}$, and the functions $A$ and $B$ and are defined as
\begin{align}\label{collect AB}
A(x,\bar x) &= {\big(E(x) - (1-x)K(x)\big)\bar K(1-\bar x)-\big( E(1-x) - xK(1-x) \big)\bar K(\bar x) \over2 \big(K(x)\bar K(1-\bar x)+  K(1-x)\bar K(\bar x)\big)}
\nn[4pt]
B(x,\bar x) &=(B(x,\bar x))^*=\bar B(\bar x, x)={\pi\over 4\big(K(x)\bar K(1-\bar x)+  K(1-x)\bar K(\bar x)\big)}
\nn[5pt]
\bar A(\bar x, x) &= ( A( x, \bar x))^*\nn[2pt]
&=
{\big(\bar E(\bar x) - (1-\bar x)\bar K(\bar x)\big) K(1- x)-\big( \bar E(1-\bar x) - \bar x\bar K(1-\bar x) \big)K(x) \over2 \big(K(x)\bar K(1-\bar x)+  K(1-x)\bar K(\bar x)\big)}
\end{align}
where $K(y)$ and $E(y)$ are the complete elliptic integrals of the first and second kinds respectively.
The cross ratio is given by (\ref{collect AB}).
\begin{align} 
x&={(z_1-z_2)(z_3-z_4)\over(z_1-z_3)(z_2-z_4)},\quad
\bar x={(\bar z_1-\bar z_2)(\bar z_3-\bar z_4)\over(\bar z_1-\bar z_3)(\bar z_2-\bar z_4)},\quad z_j=e^{w_j}, \quad \bar z_j=e^{\bar w_j}
\end{align}
where $j = 1,2,3,4$. In the following, we will perform complex conjugation. In practice, this means replacing all $z_i$ with $\bar{z}_i$ and vice versa, as well as using the complex conjugates of $A(x,\bar x)$ and $B(x,\bar x)$ from (\ref{collect AB}).

In Euclidean signature, the cylinder coordinates $w_j$ are given by
\begin{align} 
w_j &= \tau_j + i\sigma_j,\quad \bar w_j=\tau_j - i\sigma_j
\end{align}
while in Lorentzian signature
\begin{align} \label{Ls}
w_j &= i(t_j + \sigma_j),\quad \bar w_j =i(t_j -  \sigma_j)
\end{align}

\subsection*{Pair creation}

The left-moving pair creation coefficient is given by
\begin{align} \label{gamma final collect}
\gamma_{m,n} 
=\ & {1\over 4mn}(-1)^{m+n}\sum_{k_1=0}^{n-1}~\sum_{k_2=0}^{n-(k_1+1)}~\sum_{k_3=0}^{n-(k_1+k_2+1)}~\sum_{k_4=0}^{n-(k_1+k_2+k_3+1)}~\sum_{k'_1=0}^{m+n - (k_1+k_2+k_3+k_4)}\nn 
&~\sum_{k'_2=0}^{m+n - (k_1+k_2+k_3+k_4+k'_1)}\sum_{k'_3=0}^{m+n - (k_1+k_2+k_3+k_4+k'_1+k'_2)}\big(n - (k_1 + k_2 + k_3 + k_4)\big)\nn 
&{}^{{1\over2}}C_{k_1}{}^{-{1\over2}}C_{k_2}{}^{{1\over2}}C_{k_3}{}^{-{1\over2}}C_{k_4}{}^{-{1\over2}}C_{k'_1}{}^{{1\over2}}C_{k'_2}{}^{-{1\over2}}C_{k'_3}{}^{{1\over2}}C_{m+n-(k_1+k'_1+k_2+k'_2+k_3+k'_3+k_4)}\nn 
&\times\Big(z^{k_1+k'_1}_{1}z^{k_2+k'_2}_{2}z^{k_3+k'_3}_{3}z^{m+n-(k_1+k'_1+k_2+k'_2+k_3+k'_3)}_{4}\nn
&\qquad + z^{k_1+k'_1}_{2}z^{k_2+k'_2}_{1}z^{k_3+k'_3}_{4}z^{m+n-(k_1+k'_1+k_2+k'_2+k_3+k'_3)}_{3}\Big) 
\nn [3pt]
& \! - A(x,\bar x)( z_1- z_3)( z_2 - z_4)\nn 
&\quad\times\bigg({1\over \sqrt2m}(-1)^{m}\sum_{k'_1=0}^{m-1}~\sum_{k'_2=0}^{m-(k'_1+1)}~\sum_{k'_3=0}^{m-(k'_1+k'_2+1)}{}^{-{1\over2}}C_{k'_1}{}^{-{1\over2}}C_{k'_2}{}^{-{1\over2}}C_{k'_3}{}^{-{1\over2}}C_{m-(k'_1+k'_2+k'_3+1)}
\nn
&\qquad\quad z_1^{k'_1}z_2^{k'_2}z_3^{k'_3}z_4^{m-(k'_1+k'_2+k'_3+1)}\bigg)
\nn
&\quad\times\bigg({1\over \sqrt2n}(-1)^{n}\sum_{k_1=0}^{n-1}~\sum_{k_2=0}^{n-(k_1+1)}~\sum_{k_3=0}^{n-(k_1+k_2+1)}{}^{-{1\over2}}C_{k_1}{}^{-{1\over2}}C_{k_2}{}^{-{1\over2}}C_{k_3}{}^{-{1\over2}}C_{n-(k_1+k_2+k_3+1)}
\nn
&\qquad\quad z_1^{k_1}z_2^{k_2}z_3^{k_3}z_4^{n-(k_1+k_2+k_3+1)}\bigg) 
\end{align}
The left-right mixed pair creation coefficient is
\begin{align}\label{beta final}
\beta_{m,n} &= B(x,\bar x)|z_1-z_3||z_2-z_4|
\nn
&\quad\times\bigg({1\over m}(-1)^{m}\sum_{k'_1=0}^{m-1}~\sum_{k'_2=0}^{m-(k'_1+1)}~\sum_{k'_3=0}^{m-(k'_1+k'_2+1)}{}^{-{1\over2}}C_{k'_1}{}^{-{1\over2}}C_{k'_2}{}^{-{1\over2}}C'_{k_3}{}^{-{1\over2}}C_{m-(k'_1+k'_2+k'_3+1)}
\nn
&\quad\qquad~~ z_1^{k'_1}z_2^{k'_2}z_3^{k'_3}z_4^{m-(k'_1+k'_2+k'_3+1)}\bigg)
\nn 
&\quad\times\bigg({1\over n}(-1)^{n}\sum_{k_1=0}^{n-1}~\sum_{k_2=0}^{n-(k_1+1)}~\sum_{k_3=0}^{n-(k_1+k_2+1)}{}^{-{1\over2}}C_{k_1}{}^{-{1\over2}}C_{k_2}{}^{-{1\over2}}C_{k_3}{}^{-{1\over2}}C_{n-(k_1+k_2+k_3+1)}
\nn
&\quad\qquad~~ \bar z_1^{k_1}\bar z_2^{k_2}\bar z_3^{k_3}\bar z_4^{n-(k_1+k_2+k_3+1)}\bigg)
\end{align}
The right-moving pair creation coefficient is the complex conjugate of the left-moving pair creation coefficient
\begin{align} \label{gamma final prime}
\bar \gamma_{m,n} =\ & (\gamma_{m,n})^*  
\end{align}
 
\subsection*{Propagation}

The left-moving propagation coefficient is given by
\begin{align}\label{f aa collect}
f^{\alpha\alpha}_{n,p} =\, & {1\over 2p}(-1)^{p+n}\sum_{k'_1=0}^{n-1}\sum_{k'_2=0}^{n-k'_1-1}\sum_{k'_3=0}^{n-k'_1-k'_2-1}\sum_{k'_4=\max[0,n-p-k'_1-k'_2-k'_3]}^{n-k'_1-k'_2-k'_3-1}\nn 
&\sum_{k_1=0}^{p-n+k'_1+k'_2+k'_3+k'_4}~\sum_{k_2=0}^{p-n-k_1+k'_1+k'_2+k'_3+k'_4}~\sum_{k_3=0}^{p-n-k_1-k_2+k'_1+k'_2+k'_3+k'_4}(n-k'_1-k'_2-k'_3-k'_4)\nn 
&{}^{-{1\over2}}C_{k'_1}{}^{{1\over2}}C_{k'_2}{}^{-{1\over2}}C_{k'_3}{}^{{1\over2}}C_{k'_4}{}^{{1\over2}}C_{k_1}{}^{-{1\over2}}C_{k_2}{}^{{1\over2}}C_{k_3}{}^{-{1\over2}}C_{p-n-k_1-k_2-k_3+k'_1+k'_2+k'_3+k'_4}\nn 
&~~\times\bigg(\Big({z_2z_4\over z_1z_3}\Big)^{1\over2}z^{k_1-k'_1}_{1}z^{k_2-k'_2}_{2}z^{k_3-k'_3}_{3}z^{p-n-k_1-k_2-k_3+k'_1+k'_2+k'_3}_{4}\nn 
&~~~~~~+ \Big({z_1z_3\over z_2z_4}\Big)^{1\over2} z^{k_1-k'_1}_{2}z^{k_2-k'_2}_{1}z^{k_3-k'_3}_{4}z^{p-n-k_1-k_2-k_3+k'_1+k'_2+k'_3}_{3} \bigg)
\nn 
&- {A(x,\bar x)(z_1-z_3)(z_2-z_4)\over \sqrt{z_1z_2z_3z_4}}\nn 
&~~\times\bigg({(-1)^p\over p}\sum_{k_1=0}^{p-1}\sum_{k_2=0}^{p-k_1-1}\sum_{k_3=0}^{p-k_1-k_2-1}{}^{-{1\over2}}C_{k_1}{}^{-{1\over2}}C_{k_2}{}^{-{1\over2}}C_{k_3}{}^{-{1\over2}}C_{p-k_1-k_2-k_3-1}\nn
&\qquad z_1^{k_1}z_2^{k_2}z_3^{k_3}z_4^{p-k_1-k_2-k_3-1}\bigg)\nn
&~~\times\bigg((-1)^n\sum_{k'_1=0}^{n-1}\sum_{k'_2=0}^{n-k'_1-1}\sum_{k'_3=0}^{n-k'_1-k'_2-1}{}^{-{1\over2}}C_{k'_1}{}^{-{1\over2}}C_{k'_2}{}^{-{1\over2}}C_{k'_3}{}^{-{1\over2}}C_{n-k'_1-k'_2-k'_3-1}\nn
&\qquad z_1^{-k'_1}z_2^{-k'_2}z_3^{-k'_3}z_4^{-n+k'_1+k'_2+k'_3+1} \bigg) 
\end{align}
The right-moving propagation is given by the complex conjugation
\be\label{f fstar}
f^{\bar\alpha\bar\alpha}_{n,p} = (f^{\alpha\alpha}_{n,p})^*
\ee

\b

The left-right mixed propagation coefficient is given by
\begin{align}\label{f aabar collect}
f^{\alpha\bar\alpha}_{n,p} =\, &  {B(x,\bar x)|z_1-z_3||z_2-z_4| \over \sqrt{z_1z_2z_3z_4}}\nn 
&\times\bigg({(-1)^p\over p}\sum_{k_1=0}^{p-1}\sum_{k_2=0}^{p-k_1-1}\sum_{k_3=0}^{p-k_1-k_2-1}{}^{-{1\over2}}C_{k_1}{}^{-{1\over2}}C_{k_2}{}^{-{1\over2}}C_{k_3}{}^{-{1\over2}}C_{p-k_1-k_2-k_3-1}\nn
&\qquad\bar z_1^{k_1}\bar z_2^{k_2}\bar z_3^{k_3}\bar z_4^{p-k_1-k_2-k_3-1}\bigg)\nn
&\times\bigg((-1)^n\sum_{k'_1=0}^{n-1}\sum_{k'_2=0}^{n-k'_1-1}\sum_{k'_3=0}^{n-k'_1-k'_2-1}{}^{-{1\over2}}C_{k'_1}{}^{-{1\over2}}C_{k'_2}{}^{-{1\over2}}C_{k'_3}{}^{-{1\over2}}C_{n-k'_1-k'_2-k'_3-1}\nn
&\qquad z_1^{-k'_1}z_2^{-k'_2}z_3^{-k'_3}z_4^{-n+k'_1+k'_2+k'_3+1} \bigg)
\end{align}
and
\be\label{faabar faabarstar}
f^{\bar\alpha\alpha}_{n,p} = (f^{\alpha\bar\alpha}_{n,p})^*
\ee

\subsection*{Contraction}

The left-moving contraction is given by
\begin{align}\label{Caa collect}
C^{\alpha\alpha}[m,n]=\, &{1\over2}(-1)^{m+n}\sum_{k'_1=0}^{n-1}~\sum_{k'_2=0}^{n-k'_1-1}~\sum_{k'_3=0}^{n-k'_1-k'_2-1}~\sum_{k'_4=0}^{n-k'_1-k'_2-k'_3-1}~\sum_{k_1=0}^{m+n-k'_1-k'_2-k'_3-k'_4}\nn
&\sum_{k_2=0}^{m+n-k'_1-k'_2-k'_3-k'_4-k_1}~\sum_{k_3=0}^{m+n-k'_1-k'_2-k'_3-k'_4-k_1-k_2}(n-k'_1-k'_2-k'_3-k'_4)\nn 
&{}^{-{1\over2}}C_{k'_1}{}^{{1\over2}}C_{k'_2}{}^{-{1\over2}}C_{k'_3}{}^{{1\over2}}C_{k'_4}{}^{{1\over2}}C_{k_1}{}^{-{1\over2}}C_{k_2}{}^{{1\over2}}C_{k_3}{}^{-{1\over2}}C_{m+n-k_1-k_2-k_3-k'_1-k'_2-k'_3-k'_4}
\nn
&~\times\Big(z^{-k_1-k'_1}_1z^{-k_2-k'_2}_2z^{-k_3-k'_3}_3z^{-m-n+k_1+k_2+k_3+k'_1+k'_2+k'_3}_4\nn 
&\qquad + z^{-k_1-k'_1}_2z^{-k_2-k'_2}_1z^{-k_3-k'_3}_4z^{-m-n+k_1+k_2+k_3+k'_1+k'_2+k'_3}_3\Big)
\nn 
&-{A(x,\bar x)(z_1-z_3)(z_2-z_4)\over z_1z_2z_3z_4}\nn 
&~~\times\bigg((-1)^m\sum_{k_1=0}^{m-1}\sum_{k_2=0}^{m-k_1-1}\sum_{k_3=0}^{m-k_1-k_2-1}{}^{-{1\over2}}C_{k_1}{}^{-{1\over2}}C_{k_2}{}^{-{1\over2}}C_{k_3}{}^{-{1\over2}}C_{m-k_1-k_2-k_3-1}\nn 
&~\qquad z_1^{-k_1}z_2^{-k_2}z_3^{-k_3}z_4^{-m+k_1+k_2+k_3+1}\bigg)
\nn 
&~~\times\bigg((-1)^n\sum_{k'_1=0}^{n-1}\sum_{k'_2=0}^{n-k'_1-1}\sum_{k'_3=0}^{n-k'_1-k'_2-1}{}^{-{1\over2}}C_{k'_1}{}^{-{1\over2}}C_{k'_2}{}^{-{1\over2}}C_{k'_3}{}^{-{1\over2}}C_{n-k'_1-k'_2-k'_3-1}\nn
&~\qquad z_1^{-k'_1}z_2^{-k'_2}z_3^{-k'_3}z_4^{-n+k'_1+k'_2+k'_3+1}\bigg)
\end{align}
The right-moving contraction is given by the complex conjugation
\be
C^{\bar\alpha\bar\alpha}[m,n] = (C^{\alpha\alpha}[m,n])^*
\ee

\b

The left-right mixed contraction is
\begin{align}
C^{\alpha\bar\alpha}[m,n]\label{Caabar final}
=\, &{B(x,\bar x)|z_1-z_3||z_2-z_4|\over|z_1z_2z_3z_4|}
\nn
&\times\bigg((-1)^m\sum_{k_1=0}^{m-1}\sum_{k_2=0}^{m-k_1-1}\sum_{k_3=0}^{m-k_1-k_2-1}{}^{-{1\over2}}C_{k_1}{}^{-{1\over2}}C_{k_2}{}^{-{1\over2}}C_{k_3}{}^{-{1\over2}}C_{m-k_1-k_2-k_3-1}\nn 
&\qquad z_1^{-k_1}z_2^{-k_2}z_3^{-k_3}z_4^{-m+k_1+k_2+k_3+1}\bigg)
\nn 
&\times\bigg((-1)^n\sum_{k'_1=0}^{n-1}\sum_{k'_2=0}^{n-k'_1-1}\sum_{k'_3=0}^{n-k'_1-k'_2-1}{}^{-{1\over2}}C_{k'_1}{}^{-{1\over2}}C_{k'_2}{}^{-{1\over2}}C_{k'_3}{}^{-{1\over2}}C_{n-k'_1-k'_2-k'_3-1}\nn
&\qquad \bar z_1^{-k'_1}\bar z_2^{-k'_2}\bar z_3^{-k'_3}\bar z_4^{-n+k'_1+k'_2+k'_3+1}\bigg)
\end{align}
and 
\be
C^{\bar \alpha\alpha}[m,n] = (C^{ \alpha\bar \alpha}[m,n])^* = C^{ \alpha\bar \alpha}[n,m]
\ee

\section{Discussion and Conclusion}\label{discussion}

In this paper, we investigate the effect of four twist-2 operators on excited states, extending the results of \cite{Guo:2024edj}, which focused on their effect on the vacuum. We consider a simple setup where these operators act on the untwisted sector of two copies of a free boson. 
As shown in \cite{Guo:2024edj}, the effect on the vacuum is captured by a set of coefficients called pair creation, which describes the production of two modes in the final state. In this paper, we find that the effect on excited states requires two additional sets of coefficients: Propagation, which describes how one mode transforms into another, and Contraction, which describes the annihilation of two initial modes.
A novel feature discovered in \cite{Guo:2024edj} is that pair creation involves mixing between the left- and right-moving sectors. We show that this mixing also appears in the effects on excited states, meaning that propagation and contraction also exhibit left-right mixing.

To compute these effects, a six-point function involving two bosonic fields and four twist operators is required. The results are provided in \cite{Dixon:1986qv} for specific locations of the twist operators and in \cite{Guo:2024edj} for general locations. These results can be obtained using a covering map, which maps the six-point function into a two-point function of the free boson on a torus. 
This torus two-point function naturally exhibits left-right mixing, which becomes the mixing in the effect of the four twist-2 operators. For higher twists and more twist operators, this mixing is generally present whenever the covering map is not a sphere.

The ultimate goal is to understand the effect of multiple twist operators. To do this, we examine two limits of the effect of four twist operators to see if it can be simplified and reproduced by combining the effects of fewer twist operators. First, we look at the continuum limit, where the energy of the modes is large. In this case, we find that the nontrivial monodromy terms decay much faster compared to the other terms, and the mixing between left- and right-moving sectors becomes negligible. This supports the idea that, at high energies, the effect of multiple twist operators can be reproduced by combining the effects of a single twist operator \cite{Carson:2017byr}. This is because the effect of one twist operator does not involve left-right mixing, and combining them in this way will not introduce such mixing. It would be valuable to explore this further and compare it with the work in \cite{Carson:2017byr} to better understand the effect of multiple twist operators in the continuum limit. 

The second limit we consider is where the first and second twists are close together, the third and fourth twists are also close together, with a large distance between the second and third twists. Additionally, the mode numbers involved are relatively low. Naively, in the limit where the separation between the second and third twists is large, only the lowest possible energy state between them contributes. Since this state is generated by the first two twists, there is no left-right mixing. Similarly, because the final state is produced from this intermediate state through two more twists, no such mixing should occur.
Therefore, at leading order in the large-distance limit, we would expect no left-right mixing. However, our analysis shows that such mixing does occur at leading order in this limit. This suggests that the nontrivial monodromy effect, and thus the left-right mixing, is a low-energy phenomenon that cannot be ignored even when the twists are far apart.

There are several directions for further exploration. First, investigating the continuum limit and comparing it with the findings in \cite{Carson:2017byr}, to gain a better understanding of the effects of multiple twist operators in the continuum limit. Second, extending the computation to supersymmetric cases, 
particularly in the D1D5 CFT, by incorporating fermions.
Third, in the D1D5 system, the twist-2 operator is part of the marginal deformation that moves the theory away from the free point towards the supergravity regime in moduli space. Understanding the role of this left-right mixing in this process would be interesting.

\section*{Acknowledgements} 
We would like to thank Emil Martinec and Samir D. Mathur for helpful discussion and suggestions. SDH would also like to thank APCTP for their hospitality during the completion of this work. The work of SDH is supported by KIAS Grant PG096301.

\appendix
\section{Bogoliubov transformation with left-right mixing}\label{simple model}

The Bogoliubov transformation is a linear transformation that mixes creation and annihilation operators. For a simple example, see section 4 of \cite{Guo:2022sos}. In this appendix, we extend this example to include left-right mixing to help readers better understand the new effects explored in this paper.

To mimic the behavior of left- and right-moving bosons, we introduce two free bosons $a$ and $\bar a$, each has its own set of creation and annihilation operators. 
These operators satisfy the following commutation relations
\bea \label{a comm}
&&[a,a^{\dagger}]=1,\quad [\bar a,\bar a^{\dagger}]=1,
\nn 
&& [a,a]=[a,\bar a]= [\bar a,\bar a]
=[a,\bar a^{\dagger}]=0
\eea
The corresponding A-vacuum is defined by
\be \label{vacuum}
a|0\rangle_A=0,\quad \bar a|0\rangle_A=0
\ee

Now, consider a linear transformation that mixes the creation and annihilation operators
\begin{align}\label{a and b}
a &= \alpha_1 b + \alpha_2 b^{\dagger} + \alpha_3 \bar b + \alpha_4 \bar b^{\dagger}\nn 
a^{\dagger} &= \alpha_1^* b^{\dagger} + \alpha_2^* b + \alpha_3^* \bar b^{\dagger} + \alpha_4^* \bar b\nn 
\bar a &=\bar\alpha_1 b + \bar\alpha_2 b^{\dagger} + \bar\alpha_3\bar b + \bar\alpha_4\bar b^{\dagger}\nn 
\bar a^{\dagger} &=\bar\alpha^*_1 b^{\dagger} + \bar\alpha^*_2 b + \bar\alpha^*_3 \bar b^{\dagger} + \bar\alpha^*_4\bar b
\end{align}
The new set of creation and annihilation operators satisfies the following commutation relations
\bea \label{b comm}
&&[b,b^{\dagger}]=1,\quad [\bar b,\bar b^{\dagger}]=1,
\nn 
&& [b,b]=[b,\bar b]= [\bar b,\bar b]
=[b,\bar b^{\dagger}]=0
\eea
These impose constraints on the transformation parameters given by
\begin{align}\label{constraints} 
&|\a_1|^2 - |\a_2|^2 + |\a_3|^2-|\a_4|^2=1
\nn 
&|\bar\a_1|^2 - |\bar\a_2|^2 + |\bar\a_3|^2-|\bar\a_4|^2=1
\nn
&\a_1\bar\a_2-\a_2\bar\a_1+\a_3\bar\a_4-\a_4\bar\a_3=0
\nn
&\a_1\bar\a_1^* -\a_2\bar\a_2^* + \a_3\bar\a_3^*-\a_4\bar\a_4^*=0
\end{align}
The corresponding B-vacuum is defined by
\bea 
b|0\rangle_B=0,\quad \bar b|0\rangle_B=0
\eea
This B-vacuum differs from the A-vacuum. To find the relation between them, we apply the linear transformations (\ref{a and b}) to the A-vacuum condition (\ref{vacuum}), which gives
\begin{align}
( \alpha_1 b + \alpha_2 b^{\dagger} + \alpha_3 \bar b + \alpha_4 \bar b^{\dagger})|0\rangle_A&=0
\nn
( \bar\alpha_1 b + \bar\alpha_2 b^{\dagger} + \bar\alpha_3 \bar b + \bar\alpha_4 \bar b^{\dagger})|0\rangle_A&=0
\end{align} 
The solution to these equations is
\bea\label{pc}
|0\rangle_A = e^{{\gamma\over2}b^{\dagger}b^{\dagger} + \beta b^{\dagger}\bar b^{\dagger} + {\delta\over2}\bar b^{\dagger}\bar b^{\dagger}}|0\rangle_B
\eea
where the coefficients satisfy
\begin{align}
\a_1\gamma+ \a_2 + \a_3\beta&=0,& \a_1\beta+\a_3\d + \a_4&=0\nn
\bar\a_1\gamma + \bar\a_2 + \bar\a_3\beta&=0,& \bar\a_1\beta+\bar\a_3\d + \bar\a_4&=0
\end{align} 
This result illustrates that the A-vacuum, defined with respect to $a,\bar a$, is not the B-vacuum, defined with respect to $b,\bar b$. 

Instead, the A-vacuum appears as an excited state containing pairs of 
$b$ and $\bar b$ bosons from the perspective of the 
B-vacuum. This phenomenon is analogous to the pair creation effect associated with twist operators.
In this analogy, the B-vacuum represents the state before the application of the twist operator, while the A-vacuum corresponds to the state after its application. 
As evident from (\ref{pc}), the linear transformation that mixes the left and right moving bosons also induces pair creation involving these modes. This is the left-right mixing in the ansatz (\ref{pair}).
In the following, we will demonstrate that this mixing also affects contractions and propagations.

Let's now consider an excitation above the vacuum. Using the commutation relations (\ref{b comm}), we obtain
\begin{align}
a^{\dagger}|0\rangle_A &=[ (\alpha^*_2\gamma+\alpha^*_4\beta)b^{\dagger}+(\alpha^*_2\beta+\alpha^*_4\delta)\bar b^{\dagger}]|0\rangle_A\nn
\bar a^{\dagger}|0\rangle_A &=[ (\bar\alpha^*_2\gamma+\bar\alpha^*_4\beta)b^{\dagger}+(\bar\alpha^*_2\beta+\bar\alpha^*_4\delta)\bar b^{\dagger}]|0\rangle_A
\end{align}
This result shows that an a-excitation (left moving) can give rise to both a b-excitation (left moving) and a $\bar b$-excitation (right moving), and similarly for a $\bar a$-excitation (right moving). These propagation coefficients can be identified as
\begin{align} 
f^{a^{\dagger}b^{\dagger}}&\equiv \alpha^*_2\gamma+\alpha^*_4\beta,& f^{a^{\dagger}\bar b^{\dagger}}&\equiv \alpha^*_2\beta+\alpha^*_4\delta
\nn 
f^{\bar a^{\dagger} b^{\dagger}}&\equiv \bar\alpha^*_2\gamma+\bar\alpha^*_4\beta,&f^{\bar a^{\dagger} \bar b^{\dagger}}&\equiv\bar\alpha^*_2\beta+\bar\alpha^*_4\delta
\end{align}
which allows us to write
\begin{align}\label{app pc}
 a^{\dagger}|0\rangle_A &=( f^{a^{\dagger}b^{\dagger}}b^{\dagger}+f^{a^{\dagger}\bar b^{\dagger}}\bar b^{\dagger})|0\rangle_A\nn
\bar a^{\dagger}|0\rangle_A &=( f^{\bar a^{\dagger}b^{\dagger}}b^{\dagger}+f^{\bar a^{\dagger}\bar b^{\dagger}}\bar b^{\dagger})|0\rangle_A
\end{align}
Next, consider a state with two excitations
\begin{align}
a^{\dagger}\bar a^{\dagger}|0\rangle_A &= (\alpha_1^* b^{\dagger} + \alpha_2^* b + \alpha_3^* \bar b^{\dagger} + \alpha_4^* \bar b)(\bar \alpha_1^* b^{\dagger} + \bar\alpha_2^* b + \bar\alpha_3^* \bar b^{\dagger} + \bar\alpha_4^* \bar b)|0\rangle_A
\nn 
&=(\alpha_1^* b^{\dagger} + \alpha_2^* b + \alpha_3^* \bar b^{\dagger} + \alpha_4^* \bar b)(f^{\bar a^{\dagger}b^{\dagger}}b^{\dagger}+f^{\bar a^{\dagger}\bar b^{\dagger}}\bar b^{\dagger})|0\rangle_A
\nn 
&=\big((f^{a^{\dagger}b^{\dagger}}b^{\dagger}+f^{a^{\dagger}\bar b^{\dagger}}\bar b^{\dagger})(f^{\bar a^{\dagger}b^{\dagger}}b^{\dagger}+f^{\bar a^{\dagger}\bar b^{\dagger}}\bar b^{\dagger}) +C^{a^{\dagger}\bar a^{\dagger}}\big)|0\rangle_A
\end{align}
Here, in addition to the propagation terms from (\ref{app pc}), there is an extra term $C^{a^{\dagger}\bar a^{\dagger}}$, which accounts for the contraction of an $a$-excitation (left moving) and a $\bar a$-excitation (right moving), leading to their annihilation. Similar contractions can occur between two $a$-excitations or two $\bar a$-excitations.
These contraction coefficients are defined as
\begin{align}
C^{a^{\dagger}\bar a^{\dagger}} &\equiv \alpha^*_2f^{\bar a^{\dagger}b^{\dagger}}+\alpha^*_4f^{\bar a^{\dagger}\bar b^{\dagger}}\nn
C^{a^{\dagger}a^{\dagger}} &\equiv \alpha_2^*f^{a^{\dagger}b^{\dagger}}+\alpha_4^*f^{a^{\dagger}\bar b^{\dagger}}\nn
C^{\bar a^{\dagger}\bar a^{\dagger}}&\equiv \bar\alpha_2^*f^{\bar a^{\dagger}b^{\dagger}}+\bar\alpha_4^*f^{\bar a^{\dagger}\bar b^{\dagger}}
\end{align}
For more excitations, the pattern follows similarly, using the three kinds of coefficients, contraction, propagation, and pair creation as described in section \ref{results}, where these coefficients are promoted to matrices with indices corresponding to the mode numbers.

In the model discussed in section 4 of \cite{Guo:2022sos}, there is no left-right mixing, which is the case for one and two twist operators. This corresponds to taking $\alpha_3=\alpha_4=\bar\alpha_3=\bar\alpha_4=\beta=0$ in the above example and for simplicity, only focusing on the unbarred bosons.

\bibliographystyle{JHEP}
\bibliography{bibliography.bib}

\end{document}